\documentclass[12pt]{article}
\usepackage{amsmath}
\usepackage{graphicx,psfrag,epsf}
\usepackage{enumerate}
\usepackage{natbib}
\usepackage{url} 
\usepackage{amsmath}
\usepackage{amssymb}
\usepackage{lmodern}
\usepackage{amsthm}
\usepackage[dvipsnames]{xcolor}
\nonstopmode
\usepackage{url}
\usepackage{calc}
\usepackage{textcomp}
\usepackage{rotating}
\usepackage{setspace}
\usepackage{amsmath}
\usepackage{here}
\usepackage{latexsym}           
\usepackage{color}
\usepackage{graphicx}
\usepackage{here}
\usepackage{verbatim}
\usepackage[normal]{subfigure}
\usepackage{natbib}

\definecolor{mygray}{gray}{0.0}

\newtheorem{theorem}{Theorem}[section]

\theoremstyle{definition}

\newcount\footnotenumber
\def\numberedfootnote{%
  \global\advance\footnotenumber by 1
  \footnote{$^{1}$}
  }

\newcommand{\blind}{1}

\begin{document}

\def\spacingset#1{\renewcommand{\baselinestretch}%
{#1}\small\normalsize} \spacingset{1}


\if1\blind
{
  \title{A Bayesian nonlinear stationary model with multiple frequencies for business cycle analysis}
  \author{\textcolor{mygray}{\L{}ukasz Lenart\footnote{Department of Mathematics, Krakow University of Economics}, \L{}ukasz Kwiatkowski\footnote{Corresponding author: Department of Econometrics and Operations Research, Krakow University of Economics, ul. Rakowicka 27, 31-510 Krakow; e-mail: kwiatkol@uek.krakow.pl}  and Justyna Wróblewska\footnote{Department of Econometrics and Operations Research, Krakow University of Economics}}}
  \maketitle
} \fi

\if0\blind
{
  \bigskip
  \bigskip
  \bigskip
  \begin{center}
    {\LARGE\bf A Bayesian nonlinear stationary model with multiple frequencies for business cycle analysis}
\end{center}
  \medskip
} \fi

\bigskip
\begin{abstract}
We design a novel, nonlinear single-source-of-error model for analysis of multiple business cycles. The model's specification is intended to capture key empirical characteristics of business cycle data by allowing for simultaneous cycles of different types and lengths, as well as time-variable amplitude and phase shift. The model is shown to feature relevant theoretical properties, including stationarity and pseudo-cyclical autocovariance function, and enables a decomposition of overall cyclic fluctuations into separate frequency-specific components. We develop a Bayesian framework for estimation and inference in the model, along with an MCMC procedure for posterior sampling, combining the Gibbs sampler and the Metropolis-Hastings algorithm, suitably adapted to address encountered numerical issues. Empirical results obtained from the model applied to the Polish GDP growth rates imply co-existence of two types of economic fluctuations: the investment and inventory cycles, and support the stochastic variability of the amplitude and phase shift, also capturing some business cycle asymmetries. Finally, the Bayesian framework enables a fully probabilistic inference on the business cycle clocks and dating, which seems the most relevant approach in view of economic uncertainties.
\end{abstract}

\noindent%
{\it Keywords:}  stochastic cycle, innovations state space model, business fluctuations, business cycle clock.
\vfill

\newpage
\spacingset{1.75} 

\section{Introduction}

A large area of macroeconomic theoretical and empirical research is focused on business cycles, particularly their dating and prediction. The apparent importance and complexity of the subject has been driving a growing amount of research on alternative econometric approaches to the analysis and prediction of cyclical patterns in economics; see, among many others:
\cite{10.2307/1392364},
\cite{krolzig2013markov}, 
\cite{Zarnowitz_1992},
\cite{West1995BayesianII},
\cite{Layton_2001}, 
\cite{Bengoechea_2006}, 
\cite{Lin_2010}, 
\cite{Ng_2012}, 
\cite{Ferrara_2014}, 
\cite{Vigfusson_2014}, 
\cite{Pauwels_2014}, 
\cite{Aastveit_2016}, 
\cite{Chai_2016},
\cite{Nibbering_2018}, 
\cite{Tarassow_2018}, \cite{Bessec_2019},
\cite{Morikawa_2019}. 
Modelling business cycles is a complex task. Difficulties arise not only with the prediction itself, but already at the preceding phase of monitoring the current stage of a business cycle.

Choice of a particular modelling approach is of crucial importance for business cycle analysis. In many cases, it resorts to econometric models, using real-time data sets including macroeconomic data, market data, etc. However, it is widely accepted that the problem of monitoring and predicting business cycle fluctuations does not boil down solely to the choice of cyclically sensitive real-time data indicative of economic conditions. The statistical properties of a given tool play an equally important role. Hence, it appears conceivable that a properly constructed model may prove potent in cyclical fluctuations analysis. To that end, the model should feature appropriate theoretical properties corresponding to empirical data displaying cyclical (or "pseudo-cyclical") behaviour of economic activity. However reasonable such a logic may sound, formulating a purely econometric model meeting the postulate above still poses quite a challenge, lying at the very heart of this article. In particular, only recently has it been noticed (see \cite{lenart2024properties}) that in the existing cyclical models (see \cite{Hannan1970}, \cite{Harvey1985}, \cite{HarveyTrimbur2003}, \cite{GARMA1998}, \cite{LuatiProietti2010}, \cite{Proietti2023}), the amplitude of fluctuations is restricted by a linear restriction between its mean and standard deviation, namely the former is exactly $\sqrt{\frac{\pi}{4-\pi}}$ times the latter. In view of potential empirical applications of the models (not only for economic or financial time series, but cyclical data per se), this represents an extremely strong restraint. For that matter, we argue here that consequences of this involuntarily imposed constraint manifest in empirical results obtained by \cite{GARMA1998}, \cite{koopman2008measuring}, \cite{maddanu2022modelling} (see Figure 6 therein, presenting the amplitude process), \cite{Proietti2023} (see Figure 3 therein, presenting the amplitude process), and \cite{harvey2007trends} (see Figure 9 therein).

Business cycle prediction is one of the most challenging aspects of modelling economic activity fluctuations. \cite{Hamilton_2011} discusses three main difficulties related to the real-time recession forecasting. Firstly, if economic actors were able to predict recessions accurately, the latter probably would not occur. Secondly, the revised data may differ from the data which are available in real time, which may result in incorrect real-time predictions. Thirdly, key economic relationships between economic variables change continually over time. With respect to the first issue, business cycle forecasts always display a great deal of uncertainty, however wide a range of information and available methods one uses to obtain them. The second problem is impossible to avoid (as of today) because inevitably data revisions are still in operation. To cope with the third difficulty, models with time-varying parameters are broadly used (see, e.g., \cite{Agostino_13}). On the other hand, estimation of such models is riddled with problems (see \cite{Onorant_2019}).

The main objective of the article is three-fold. Firstly, drawing upon results presented in \cite{lenart2024properties}, we construct a new class of non-Gaussian models for business cycle analysis, with their specification closely related to the well-known family of the single
source of error models (SSOE), (see, e.g., \cite{ord1997estimation}, \cite{Hyndman_08_book}). Secondly, we aim to propose and develop such models which would be flexible enough to take into account typical, well-known features of business cycles, including time-varying (rather than fixed) both amplitude and length of cycles, as well as coexisting (rather than single) cyclical components of different lengths. Finally, we also develop a Bayesian framework for the estimation and inference in the resulting stochastic cycle model, along with relevant numerical methods for posterior sampling, hinged on two 'classic' Markov Chain Monte Carlo (MCMC) methods: the random walk Metropolis-Hastings algorithm and the Gibbs sampler. 

As for the first of the paper's objectives, it is worth noting that the SSOE models, has been successfully employed to provide a valid statistical framework (maximum likelihood estimation, prediction intervals, model selection, etc). Each SSOE model comprises an observation (or measurement) equation and transition (or state) equations, all driven by the very same, single source of error. Most often, the state equations describe separately the dynamics of each: the level, trend and seasonality. There are a lot of particular specifications of SSOE models (see \cite{Hyndman_08_book} for details). However, as yet no SSOE specification that would feature the so-called \emph{cyclic pattern} has been considered in the literature.\footnote{Note that the very term of \emph{cyclic pattern} is of a rather descriptive nature, lacking rigorous and formal definition in the literature. Here, by the \emph{cyclic pattern} we mean the cyclic fluctuations which are not of a fixed frequency.} Most often, such \emph{cyclic patterns} are of economists' main interest in the context of business cycle analyses. Note that \cite{Hyndman_08_book} (p. 176) propose some basic model structure containing a \emph{cyclic pattern} and formulates it in the innovations state space approach. However, not only is this model the only of the sort that can be found in the ESM literature, but also its empirical application and verification have not been considered in the literature yet. Therefore, one could wonder about the reason behind this gap, despite the undisputed popularity of ESM. To answer it, notice that the basic problem with the construction of such a model is formulation of an appropriate equation, representing the so-called \emph{cyclic component} (next to the level, trend and seasonality components), capturing the evolution of the \emph{cyclic pattern} of the data. Apparently, this still remains an open problem, somehow having been absent from the ESM researchers and practitioners' scope of interest so far. Therefore, we aim at filling this gap by formulating relevant SSOE models with cyclic component, analysing their theoretical properties and finally, illustrating and validating them empirically. To this end, it is a prerequisite to specify a process that could be incorporated as a \emph{cyclic component} into an SSOE model. Note that not all econometric models that are typically applied in business cycle analyses can be adopted for this purpose, due to a specific construction of models formulated in the innovations state space framework. First of all, the \emph{cyclic component}, further denoted by $\{c_t\}$, of an SSOE model in the innovations state space form can (although does not have to) be covariance stationary, which is a generally accepted paradigm in the literature. Then, such a stationary component $\{c_t\}$ should be characterized by a pseudo-cyclical autocovariance function with pseudo-periods associated with a given (estimated) set of frequencies related to the lengths of the cycles. Equivalently, this property indicates that the mass under the spectral density function should have the ability to concentrate around these (estimated) frequencies. Only such stationary models can be able to replicate and extrapolate the pseudo-cyclical behaviour of economic activity. Alternatively, in relation to some results existing in the literature, this stationary component $\{c_t\}$ can be termed a \emph{stochastic cycle}. On the other hand, non-stationary models can be proposed for the \emph{cyclic component} $\{c_t\}$, such as ARMA processes with complex unit roots at given frequencies corresponding to business fluctuations (see \cite{Bierens_2001}). Summing up, the first (theoretical) goal of this article is to fill the gap in the literature by specifying a valid model that can describe cyclic patterns in economic data, and which can be used as the cyclic component in ESM.

With respect to the second aim of the paper, time-varying amplitude and length of business fluctuations for cyclical indicators are recognized as ones of basic characteristics of business cycles (see \cite{Zarnowitz_1992}); and in the more general context of modeling various cyclical data: (see \cite{paraschakis2012frequency}, \cite{marron2015functional}, \cite{proietti2017seasonal}). Moreover, it is not the lack of sufficient sample size but actually the unexpected changes in the length and amplitude of a cycle that need to be addressed. Consequently, recognizing the peaks and troughs of a business cycle or the duration of recessions or contractions (for example according to the National Bureau of Economic Research definition advanced in a seminal paper by \cite{Burns_46}) poses quite a challenge. Thus, in this article, we develop such business cycle models that would feature time-varying amplitude and cycle length, rendered so, however, in such a manner that relieves the models from a ubiquitous Gaussianity and the concomitant restraint between the mean and standard deviation of the amplitude. To that end, first, we utilize the idea of time-varying components in the innovations state space framework, and second, specify \textit{directly} some suitable stochastic processes that govern the amplitude and phase shifts. Interestingly, our approach makes do without more than a single process driving the amplitude, regardless of the number of frequencies, which runs counter the extant literature on multiple frequencies models, where, for purely theoretical reasons, mutually independent amplitudes are considered for the frequencies (see \cite{lenart2024properties}). However reductive our approach may appear at first glance (as hinged upon a single, common amplitude instead of separate, frequency-specific and independent amplitudes), we show that our models' specification liberates us from the 'curse of Gaussianity' (still afflicting the extant multiple frequencies models), thereby largely facilitating our models' flexibility in capturing the amplitude. \cite{Beaudry_Galizia_Portier_2016_Putting_the_Cycle_Back_into_Business_Cycle_Analysis} estimated and examined spectral density functions for several cyclically sensitive variables, and pointed to their pronounced multimodality. In relation to modelling business cycle fluctuations, the motivation to develop a \emph{cyclic pattern} (or, \emph{stochastic cycle}) containing several coexisting cycles stems from the widely known typology of business cycles according to their periodicity (see \cite{Schumpeter_1954}). Commonly, the following four types of cycle are considered: the Kitchin inventory cycle (of 3 to 5 years), the Juglar investment cycle (of 7 to 11 years), the Kuznets infrastructural investment cycle (of 15 to 25 years), and the Kondratiev wave (or long) technological cycle (of 45 to 60 years). Since the source of each of these fluctuations is different, it sounds reasonable to specify a business cycle model in such a way that would allow for a joint coexistence of different cycles. In view of the above, it appears that there is still a lot of room for improvement in that regard from an econometric and statistical perspective, since no \emph{cyclic pattern} with many coexisting cycles (within the innovation state space approach) has been considered in the literature as yet. Multiple separate cyclical components are also possible to specify in an alternative framework of the structural time series models (see \cite{Harvey_89}), with the stochasticity of each component driven by a different error term. Moreover, in this paper we aim to fill the gap within the area of the SSOE models, where no specifications allowing for cyclical patterns have been proposed so far. To fill the gap, in this article we introduce and investigate both the theoretical and empirical properties of such business cycle models that allow for many (rather than single) coexisting cycles of time-varying (rather than fixed) amplitudes and lengths.

The paper is organized as follows. In Section 2, we present a general background behind the stochastic (as opposed to deterministic) cycle model introduced in this work, and deliver some theorems displaying theoretical properties of the model (with the proofs provided in Section C of the Supplementary Material. In Section 3, we develop a Bayesian framework for the estimation of the model, designing a relevant MCMC sampler to simulate from the posterior distribution of the parameters. Sections 4 illustrates the model estimation with on a real-world data of the quarterly Polish GDP growth rate. Section 5 summarizes and discusses some additional issues regarding the model specification.

\section{The genesis and properties of the model}

\subsection{General background}

For a univariate observable time series $\{y_t\}$ the most popular model-based approach in business cycle analysis is the state space framework, as introduced and developed by \cite{Harvey_1985}, \cite{Harvey_89}, and \cite{Harvey_Jaeger_1993}. Within this setting, a univariate cyclic component $\{c_t\}$  (or, \emph{stochastic cycle}) is assumed to be a zero-mean stationary time series and can be reduced to an ARMA($2$,$1$) process with conjugate complex roots in the autoregression part, with its characteristic polynomial 
$\Psi(L)=1-2 \rho \cos(\omega) L + \rho^2 L^2$, 
where $L$ is the lag operator, $L^k y_t = y_{t-k}$ for $k \in \mathbb{N}$,  $\omega \in  (0,\pi]$ is the frequency (measured in radians) related to the cycle length of $2\pi/\omega$, and $\rho \in [0,1]$ is a damping factor. In such a case (and under $\rho \in (0,1)$), the mass under the spectral density function has the ability to concentrate around a given frequency $\omega$, related to cyclical fluctuations. This stochastic cycle can be represented as an element of a bivariate VAR($1$) process, where the autoregressive coefficient matrix is related to the circular motion on a plane.  

For several years, this concept has been used in modelling and forecasting business cycles.It was first generalised by \cite{Harvey_Trimbur_2003} and \cite{trimbur_06}, who introduced and shown key properties of an \emph{$n^{th}$-order stochastic cycle}, including the closed form of the spectral density and autocorrelation functions. Another generalisation was proposed in \cite{Luati_Proietti_10}, where the proposed stochastic cycle is shown to reduce to an ARMA(2,~1) process, but the cyclical dynamics originates from the motion of a point along an ellipse (rather than a circle). Thereby, such a stochastic cycle specification can account for certain types of asymmetries. In the same paper, the authors generalised this model to a \emph{$n^{th}$-order stochastic cycle} with an ARMA($n$, $n-1$) form and noticed that it can allow for multiple modes in the spectral density function. However, the number and locations of peaks in the spectral density were not examined in the cited paper, probably due to a rather complicated form of the proposed model.

Presently, the concept of a stochastic cycle that can be reduced to some ARMA process is still being investigated in the literature. This concept appears to be the most popular in modelling business or financial cycles (see, e.g., \cite{koopman_azevedo_rua_06}, \cite{Harvey_07}, \cite{koopman_azevedo_06}, \cite{Koopman_Shephard_15}, among many others). However, in the context of business cycle \textit{forecasting} only the simplest version of the model proposed by \cite{Harvey_Trimbur_2003} (and further investigated in \cite{trimbur_06}), featuring $n=1$, has been considered so far (see, e.g., a recent application by \cite{Nibbering_2018}).

For most macroeconomic time series displaying cyclic pattern, the assumption of a stochastic nature of business fluctuations seems relevant, which is due to the time-varying dynamics of a business cycle. On the other hand, \cite{harvey_04x} described the idea of a \emph{fixed deterministic cycle} with one frequency (\emph{deterministic cycle}, in short) of the form: $y_t = a \sin(\lambda t +p) + \epsilon_t$, where $a \in \mathbb{R}$ is the amplitude, $\lambda \in (0,\pi)$ is the frequency (measured in radians), $p \in [0,\pi)$ is the phase shift, and $\{\epsilon_t\}$ is a white noise\footnote{\textcolor{black}{However, in \cite{TPSeasonalchanges2017}, the term `deterministic cycle' is used for the case of no noise contamination, i.e. $\epsilon_t \equiv 0$.}}. However, this alternative model has never gained as much popularity in business cycle analyses as the stochastic cycle, which apparently is attributable to a valid conjecture that a deterministic cycle is simply not flexible enough to capture the complex nature of business cycles found in empirical data. Recently, the idea of the deterministic cycle with one frequency was extended to the multiple-frequency case in  \cite{Lenart_Mazur_Pipien_16_EQ} and \cite{Lenart_Pipien_17_CEJEME}, where the existence of particular and common deterministic cycles was tested by means of a subsampling approach and the Fourier representation of almost periodic functions (see \cite{Corduneanu_1989}). Although some common deterministic cycles have indeed been identified in the above-mentioned papers, the concept of deterministic (rather than stochastic) cycles may still appear to be lacking in terms of flexibility, also in forecasting, since the conditional expected value (i.e. the point forecast) assumes the almost periodic form, which seems too strong an assumption, given the time-varying nature of business cycles for most free-market economies. However, the almost periodic functions can still be a useful foothold in construction of cycles endowed with stochasticity.

Again, business cycles are dynamic and complex phenomena, with their four alternating phases arriving on \textit{a priori} non-specified times, with their time-varying durations and amplitudes. Consequently, modelling and prediction of business cycles fluctuations, despite growing data sets, pose quite a challenge, requiring the models used to be able to adapt to the inherently changing features of a business cycle. It appears that the SSOE models can be just well-suited for the job, as long as we are able to propose a flexible enough cyclic component. As mentioned before, the state space models are broadly used in time-series analysis, also in business cycle modelling, and allow considerable flexibility in specification of the model's parametric structure. In this paper, we will use a specific formulation of a state space model, called the \emph{innovations state space}, wherein all of the error sources are perfectly correlated (see \cite{Anderson_79}, \cite{Aoki_90}, \cite{Hannan_1988}). Due to this feature, such structures are also referred to as the \emph{single-source-of-error} models (e.g., \cite{Ord_1997}). Linear innovations state space models have been successfully employed to provide a valid statistical framework for SSOE methods (see, e.g., \cite{Hyndman_08_book}). 

For an observable time series  $\{y_t\}$, a \emph{linear innovations state space model} can be written as (see \cite{Hyndman_08_book} for details): $y_t=\textbf{w}'\textbf{x}_{t-1} + \epsilon_t$ (the observation equation), $\textbf{x}_t= \textbf{F} \textbf{x}_{t-1} + \textbf{g} \epsilon_t $  (the state equation), where $\epsilon_t$ is a white noise with an unknown variance, matrices $\textbf{F}$, $\textbf{g}$ and $\textbf{w}$ comprise unknown coefficients (parameters), and $\textbf{x}_{t}$ is a \emph{state vector}. Note that the state equation can be written as $\textbf{x}_t= \textbf{F} \textbf{x}_{t-1} + \textbf{g} (y_t - \textbf{w}'\textbf{x}_{t-1})$, which indicates that $\textbf{x}_t$ is observable under the seed state vector $\textbf{x}_0$. Generally, $\textbf{x}_t$ comprises (unobserved) components that describe the evolution of the level, trend and seasonality of the series (and possibly, also a cyclic pattern component, but, as mentioned before, it has not been considered in the literature yet). In a more general setting, one can also consider a \emph{nonlinear innovations state space model}, which can be formulated as (see details in \cite{Hyndman_08_book} or \cite{Ord_1997}): $y_t=\textbf{w}(\textbf{x}_{t-1}) + \textbf{r}(\textbf{x}_{t-1}) \epsilon_t $, $ \textbf{x}_t= \textbf{f} (\textbf{x}_{t-1}) + \textbf{g}(\textbf{x}_{t-1}) \epsilon_t$, where $\textbf{w}(\cdot)$ and $\textbf{r}(\cdot)$ are some scalar functions, while $\textbf{f}(\cdot)$ and $\textbf{g}(\cdot) $ are some vector functions. The stochastic cycle model introduced in this paper admits the above nonlinear form.


\subsection{Fixed deterministic cycle model}

To facilitate the presentation of a multi-frequency stochastic cycle specification introduced in the following subsection, we briefly outline a single frequency deterministic cycle model, introduced by \cite{harvey_04x}. Upon a reparameterisation, the latter follows the equation:
\begin{equation}
y_t = a \sin(\lambda (t+p)) + \epsilon_t,
\label{deterministic_cycle}
\end{equation}
where $a \in \mathbb{R}$ denotes the amplitude, $\lambda \in (0,\pi)$ is the frequency, $p \in \left[0, \frac{\pi}{\lambda} \right]$ is the phase shift (with the support confined to ensure identification), and $\epsilon_t$ is a white noise. Note that $E(y_t) = a \sin(\lambda (t+p))$ is an almost periodic function with a single frequency $\lambda$. 

In our paper, before we allow for the stochasticity of the cyclic component, we extend model (\ref{deterministic_cycle}) into the multiple-frequency setting:  
 \begin{equation}
y_t = \underbrace{a \sum\limits_{j=1}^k  q_j \sin(\lambda_j (t+p_j))}_{f(t)} + \epsilon_t,
\label{multi_deterministic_cycle}
\end{equation}
where $q_1=1$ (for the sake of identification), $q_j \in \mathbb{R}$ for $j>1$, while $p_j \in \left[0, \frac{\pi}{\lambda_j} \right]$ is the phase shift for $j=1,2,\ldots,k$.  In such a model, the expectation $f(t)$ of $y_t$ is an almost periodic function with frequencies $\lambda_1,\lambda_2, \ldots , \lambda_k$, satisfying the restrictions $\forall_{j=1,...,k}\, \lambda_j \in \left[0, \pi \right] $ and $\lambda_i \ne \lambda_j$ for $i,j=1,...,k$, $i\ne j$. To ensure these conditions it may be convenient to impose the order constraint, for example $0\leq \lambda_1 < \lambda_2 < \ldots < \lambda_k \leq \pi$. Note that any almost periodic function $\sum_{j=1}^k A_j \sin(\lambda_j t) + B_j \cos(\lambda_j t)$ with frequencies  $0\leq \lambda_1 < \lambda_2 < \ldots < \lambda_k \leq \pi$  can be represented in terms of the above formulated function $f(t)$ and \textit{vice versa}. Therefore, $f(t)$ represents the class of almost periodic functions with a finite set of frequencies.

\subsection{Stochastic cycle model -- construction and properties}

Model (\ref{multi_deterministic_cycle}) assumes constant parameters of the amplitude and phase shifts, which is a rather strong restriction in view of the features exhibited by real-world cyclical data. Therefore, in this paper, we further extend it by allowing for time-variability of the parameter $a$ and phase shifts $p_1,p_2,\ldots,p_k$. To that end, we resort to the idea of the innovations state space models (see \cite{Hyndman_08_book}), with the resulting model admitting the following form:
\begin{equation}\left\{
\begin{array}{rll}
  y_t =& (a+A_{t-1}) \sum\limits_{j=1}^k q_j \sin(\lambda_j (t+p_j+P_{t-1})) + \mu(t) + \epsilon_t & \text{ }\\
  \Phi(L)A_t =&  \alpha_A \epsilon_t  \,\,\,\,\,\,\,\,\,\,\,\,\,\,\,\,\,\,\,\,\,\,\,\,\,\,\,\,\,\,\,\,\,\,\,\,\,\,\,\,\,\,\,\,\,\,\,\,\,\,\,\,\,\,\,\,\,\,\,\,\,\,\,\,\,\,\,\,\,\,\,\,\,\,\,\,\,\,\,\, \text{   deviations from amplitude } a &  \\
  P_t =&  \psi_P P_{t-1} + \alpha_P \epsilon_t \,\,\,\,\,\,\,\,\,\,\,\,\,\,\,\,\,\,\,\,\,\,\,\,\,\,\,\,\,\,\,\,\,\,\,\,\,\,\,\,\,\,\,\,\,\,\,\,\,\,\,\,\,\,\, \text{ stochastic phase shift} & 
\end{array}
\right.
\label{model_root}
\end{equation}
where $A_t$ and $P_t$ follow the $p^{th}$- and first-order autoregressions, respectively, $\alpha_A, \alpha_P \in \mathbb{R}$, and $\psi_P \in (-1,1]$. The polynomial  $\Phi(L)=1-\phi_1 L -\ldots - \phi_p L^p$  has roots outside the unit circle (to ensure the stationarity of $A_t$), and $\epsilon_t$ follows a Gaussian white noise with variance $\sigma^2$. Note that the very same innovations govern the dynamics of $y_t$, $A_t$ and $P_t$, which is typical to the innovations state space approach (see \cite{Hyndman_08_book}). Possible deterministic components (a constant, trend) are placed in $\mu(t)$. Some very preliminary results concerning the above model in the special case of only one frequency (and only under $p=1$) can be found in  \cite{Lenart_Wroblewska_18}. In the current paper, we generalise the approach presented in the cited work by allowing for multiple frequencies and also for any $p \in \mathbb{N}$.

Adding a stochastic component \textcolor{black}{$A_t$} to the parameter $a$ (see the first equation in model \ref{model_root}) is meant to capture possible time-variability of the overall amplitude of the cyclical fluctuations, while $P_t$ is related to a time-varying length of the cycle. To capture possible changes in the phase shifts related to each sine component ($j=1, 2, ..., k$), to each constant component $p_j$ the same stochastic process $P_t$ is introduced. Although allowing for separate processes $P_t^{(j)}$ for each $j$ could appear conceptually appealing, we feel that such an extension might result in over-parameterisation of the model, hindering estimation of its parameters. Nevertheless, such considerations are beyond the scope of the current paper and thus deferred for future research.

To recap, the overall amplitude of the process is driven by the three: the constant component $a$, a stationary $A_t \sim AR(p)$ process of amplitude fluctuations, and frequency-specific $q_j$s coefficients, $j=1,2, ..., k$. Finally, the phase shifts hinge upon frequency-specific constants $p_j$s, each shifted by the same random walk $P_t$.

Notice that the stochasticity of the amplitude and phase shifts has already been introduced in the literature, yet in a different manner. The genesis of the most popular stochastic cyclical model is based on the motion of a point along a circle, with the motion being controlled by Givens rotation matrix $\boldsymbol{R}(\lambda)$ (see, for example, \cite{west1989bayesian} and \cite{Pelagatti_16}). Rescalling this matrix with a damping parameter $\rho$ and introducing an additive error term results in a stationary stochastic cycle with a restricted ARMA(2,~1) form. As seen in (\ref{model_root}), in this paper we use quite a different and novel idea to obtain a stationary stochastic cycle. Rather than employing a damping parameter, we introduce stochastic variability in both the phase shift of the sinusoids as well as their amplitude.

As assumed above, $A_t$ is covariance stationary, as it seems plausible that typically the amplitude of business cycles would not exhibit non-stationary movements. On the other hand, for $P_t$ we also admit the case of a non-stationary, random walk process ($\psi_P = 1$). To justify such a possibility we first present two major theoretical results obtained for $Y_t$ under non-stationary and, separately, stationary $P_t$. To that end, let us introduce some notation: $\textbf{g}=[1\,\,\,0\,\,\,0\,\,\,\ldots\,\,\,0]'$ be a vector of length $p$, with $(\cdot)'$ denoting the vector transposition, and let $\textbf{F}=[f_{ij}]$ be a $p\times p$ matrix with elements $f_{1j}=\psi_{j}$  and $f_{i,(i-1)}=1$ for $i,j \in \{1,2,\ldots,p\}$. Finally, $\textbf{I}$ denotes the identity matrix. The following theorem (proof in Section C of the Supplementary Material presents essential stochastic properties of $Y_t$ under $P_t$ following a random walk ($\psi_P = 1$).

\begin{theorem}
Let $\gamma_{A}(\tau)$ denote the autocovariance function of $A_t$, i.e. $\gamma_{A}(\tau)=\text{Cov}(A_t,A_{t+\tau})$, $\tau \in \mathbb{Z}$.  If  $\psi_P=1$, then $y_{t}-\mu(t)$ is a zero-mean stationary time series with autocovariance function $\gamma_y(\tau)=\text{Cov}(y_t,y_{t+\tau})$ of the form:
$$\gamma_{y}(\tau) \!=\! \frac{1}{2} \!\sum\limits_{j=1}^k \!  q_j^2  e^{-\frac{|\tau| \alpha_P^2 \lambda_j^2 \sigma^2}{2}}
\!\!\!\cos(\lambda_j |\tau|)( \gamma_{A}(|\tau|) \!+\!a^2)
\sin( \lambda_j |\tau|) a \lambda_j \alpha_A \alpha_P  \sigma^2 \emph{\textbf{g}}' (\emph{\textbf{I}}\!-\!\emph{\textbf{F}}^{|\tau|})(\emph{\textbf{I}}\!-\!\emph{\textbf{F}})^{-1}\! \emph{\textbf{g}}
$$
for $|\tau|>0$, and 
\begin{equation}
\text{Var}(y_{t})=\gamma_{y}(0)=    \frac{1}{2}\left(\sum\limits_{j=1}^k  q_j^2 \right)(\gamma_A(0)+a^2) + \sigma^2.
\label{eq_var}
\end{equation}
\label{psi_T=1_ARp}
\end{theorem}
\vspace{-0.0 cm}
In the \emph{$n^{th}$-order stochastic cycle} model developed by \cite{Harvey_Trimbur_2003} and \cite{trimbur_06}, the autocovariance function at lag $\tau$ is proportional to $\cos(\lambda \tau)$, with $\lambda$ being the frequency related to the length of the period (see Proposition 3 in \cite{trimbur_06}). This property documents the cyclical (or, synonymously, 'pseudocyclical'; see \cite{trimbur_06}) behaviour of the model. Our result presented in the theorem above also shows the explicit formula for the autocovariance function, which depends explicitly on all frequencies $\lambda_1,\lambda_2,\ldots,\lambda_k$, and is proportional to a linear combination of the terms $\cos(\lambda_j \tau)$ and $\sin(\lambda_j \tau)$. Note also that the variance of observations is driven by their conditional variance $\sigma^2$ and quantities pertaining to the amplitude ($a$, $\text{Var}(A_t)$ and $q_j$s), while remaining independent of the phase shifts components ($p_j$s); see Equation (\ref{eq_var}).

Theorem \ref{psi_T=1_ARp} states that under $P_t$ following a random walk, $y_t-\mu(t)$ is a zero-mean stationary process with a cyclical autocovariance function. Interestingly, however, one can prove that under $P_t$ defined as a stationary AR(1) (i.e. $|\psi_P|<1$), the mean function of $y_t-\mu(t)$ is a non-zero (in general) almost periodic function with frequencies $\lambda_1,\lambda_2,\ldots,\lambda_k$. We formulate this statement in the theorem below (proof in Section C of the Supplementary Material.

\begin{theorem}
If $|\psi_P|<1$, then $y_t-\mu(t)$ features an almost periodic mean function  
$E(y_t)=\mu(t) + a \sum\limits_{j=1}^k q_{j} e^{-\frac{1}{2}\lambda_j^2 \sigma_{P}^2} \left( \sin(\lambda_j (t+p_j))  + \cos(\lambda_j (t+p_j))e^{-\frac{1}{2}\lambda_j^2 \sigma_{P}^2} \sigma_{AP} \lambda_j\right)$
with frequencies $\lambda_1,\lambda_2,\ldots,\lambda_k$, where $\sigma_{AP}=\text{Cov}(A_t,P_t)$ and $\sigma_P^2=\text{Var}(P_t)=\alpha_P^2 \frac{\sigma^2}{1-\psi_P^2}$. 
\label{psi_T<1}
\end{theorem}

With respect to both of the theorems formulated above, let us emphasize two aspects. Firstly, note that the almost-periodicity of the mean function somehow brings us back to the fixed deterministic cycle setting (see  \cite{harvey_04x} and discussion in \cite{lenart2024properties}), yet only in terms of the expectation rather than the autocovariance function. Although we are not able to derive the latter for our model under $|\psi_P|<1$, one can conjecture that it would also exhibit some cyclicality, as opposed to the deterministic cycle specifications. However, such a model in which the cyclical patterns in the mean and autocovariances are intertwined (sharing the same amplitude and phase shifts parameters in a complex, nonlinear fashion) appears of little pragmatic merit, giving way to an alternative approach of simply introducing some deterministic components in $\mu(t)$ in the model with $P_t$ defined as a random walk. Therefore, in what follows, we limit our considerations only to the case of $\psi_P=1$, omitting the one of a stationary $P_t$.

Secondly, notice that the stationarity of the model hinges upon the stochastic properties of the phase shifts process, $P_t$, and not the stochasticity of the amplitude, $A_t$. In fact, the latter could be specified as deterministic, without affecting the two models' properties stated in Theorems \ref{psi_T=1_ARp} and \ref{psi_T<1}. Introducing time-variability also into the amplitude is meant to enhance the model's flexibility.

Some illustrative simulation results pertaining to the model's spectral density and autocorrelation function are discussed in Section A of the Supplementary Material.

\subsection{Relation with other existing cyclical components}

As mentioned earlier, in the existing literature the construction of cyclic models is based on relevant equality restrictions for the parameters of ARMA models. The major goal in designing a cyclic model is such specification of all of its cyclic components that the mass under the resulting spectral (or pseudo-spectral) density function  concentrates strongly around each of the corresponding frequencies related to cyclic fluctuations. To that end, \cite{West1995BayesianII} proposes to specify each cyclic component as a latent non-stationary AR(2) process with one unit root and the other one being a free root controlling the length of cyclical fluctuations (the sum of such components define the final model). The ability of strong mass concentration around given frequencies in the pseudo-spectral density function (with the density values tending to infinity at the frequencies) is owed to the presence of the unit root. Moreover, since each of the AR(2) components is defined as a latent process, it is driven by its own, component-specific error term. Compared to such an approach, the model developed in this paper is quite different, since all of its components are governed by a single innovation term, yet in a nonlinear fashion. Furthermore, instead of autoregressive processes, each cyclic component is based here directly on a sine function, explicitly incorporated in the model's equation, additionally with their amplitudes and phase shifts evolving stochastically according to a stationary AR process and random walk. As shown in the previous section, such a model, contrary to the one developed by \cite{West1995BayesianII}, is stationary (with zero mean and cyclical autocovariances), and thus seems to be more appropriate for forecasting, particularly in the mid- and long-term perspective.

Yet another approach, also hinged on ARMA structures, was developed by \cite{Harvey_Trimbur_2003}, \cite{trimbur_06}, and \cite{Luati_Proietti_10}. The mass concentration around a given frequency is obtained there by the multiplication of the same (multiple) complex roots of the characteristic polynomial related to the AR part of an ARMA structure. The model enjoys stationarity but becomes somewhat complicated in statistical inference. Our model also differs significantly from the one proposed in the aforementioned works, as already argued in the previous paragraph with respect to the model designed by \cite{West1995BayesianII}.

A very innovative and general model was proposed in \cite{dahlhaus2017statistical}, where statistical inference has been developed for a generalized state space model with a nonlinear observation equation (referred to as the `oscillation process') given by $y_t = a_t f(\phi_t) + \epsilon_t$ with a periodic function $f$, amplitude process $a_t$, phase process $\phi_t$, and noise $\epsilon_t$. However, the statistical properties (including stationarity) of such a specification have not been investigated. Noteworthy, our proposal corresponds strongly to this general model structure, with our model assuming specifically $f(\cdot)=\sin(\cdot)$ and reducing independent innovation terms that drive $y_t$ and two unobservable components: $a_t$ and $\phi_t$, to a single source of error.

In summary, the model developed in this paper ensures strong mass concentration (but without density values tending to infinity) around given frequencies by a direct use of sine functions, while the ARMA structures considered in  \cite{Harvey_Trimbur_2003}, \cite{trimbur_06}, \cite{Luati_Proietti_10},  \cite{West1995BayesianII} achieve the effect through either introducing non-stationarity or imposing strong parameter restrictions. 

\section{Bayesian estimation of the model}

To estimate the model proposed in this paper we resort to the Bayesian approach, necessarily facilitated by Markov Chain Monte Carlo (MCMC) procedures for sampling the posterior distribution. Apparently, to date, this alternative to the frequentist statistical framework has gained fairly little popularity among both researchers and practitioners in the field of SSOE methods, where the 'classic' maximum likelihood estimation still prevails. Single papers developing Bayesian framework for some uni- and multivariate SSOE models (of, however, quite simple structures) include \cite{Forecasting_correlated_time_series_with_exponential_smoothing_models_2011}, \cite{Multivariate_exponential_smoothing_A_Bayesian_forecast_approach_based_on_simulation_2009} and  \cite{Lenart_Wroblewska_18}.

Nevertheless, in our considerations we resort to the Bayesian setting for at least two reasons. Firstly, the Bayesian inference is free from sample-size asymptotics (although MCMC algorithms are founded upon 'their own' asymptotic underpinnings). The latter is of particular merit in modelling and forecasting macroeconomic data sets, for example, which usually form relatively short samples.

Secondly, the Bayesian framework allows to formally account for the model's parameters uncertainty, which is essential in view of the model's nonlinear and complex structure. Although the likelihood function of model (\ref{model_root}) can be written down quite easily (as shown below), its form is still quite intricate, and its surface riddled with multiple local maxima, making the maximization quite a challenge. Noteworthy, the multimodality of the likelihood is already encountered in the models of only one frequency (see, e.g., \cite{Bretthorst_1988}, \cite{Lenart_Mazur_16}). Thus, it may be expected that with introducing more frequencies, related to cycles of different lengths, the problem will only exacerbate. This conjecture is corroborated by the analysis presented by \cite{Bretthorst_1988}, where the posterior distribution (approximated by the periodogram of data) examined for some simple fixed deterministic cycle with only one frequency already featured multiple modes. Multiple maxima of the likelihood (and thus, the posterior distribution) reveal a more complex picture of parameters' uncertainty, which should be accounted for in both the in-sample inference and prediction. To that end, the Bayesian paradigm is an obvious choice.

Despite the apparent appeals of the Bayesian approach, it requires fairly sophisticated numerical tools (MCMC), the use of which may not be straightforward, particularly in complex models such as the ones considered here. On the other hand, in our experience it feels that it is still far more worth the effort than the endeavor of numerical maximization of a (most likely) highly multimodal likelihood function. 

We develop a Bayesian approach to the estimation of the model (\ref{model_root}) under $\psi_P=1$. Note that \cite{Lenart_Wroblewska_18} have already discussed Bayesian inference for a similar specification, yet featuring only one frequency and actually without providing empirical illustrations. The model's parameters are collected in vector $$\boldsymbol{\theta}=(A_0,A_{-1},\ldots,A_{-p+1},a,\textbf{q},\boldsymbol{\lambda},\textbf{p},\boldsymbol{\beta},\boldsymbol{\phi}, \alpha_A, \alpha_P,\omega),$$ where $(A_0,A_{-1},\ldots,A_{-p+1}) \in \mathbb{R}^p$ are the initial conditions for $A_t \sim  AR(p)$; $a \in \mathbb{R}$, $\omega = 1/\sigma^2 \in \mathbb{R}_{+}$ stands for the error term precision, $\textbf{q}=(q_2,q_3,\ldots,q_k) \in \mathbb{R}^{k-1}$ (for $k>1$); $\boldsymbol{\lambda}=(\lambda_1,\lambda_2,\ldots,\lambda_k) \in [0,\pi]^k$; $\textbf{p}=(p_1,p_2,\ldots,p_k) \in \prod_{j=1}^k [0,\frac{\pi}{\lambda_j}]$; and $\boldsymbol{\phi}=(\phi_1,\phi_2,\ldots,\phi_p) \in \textbf{C}_{p}$, with $\textbf{C}_{p} \subset \mathbb{R}^p$ denoting the stability (or, stationarity) region. The initial conditions for $A_t$ are explicitly modelled here, while the one for $P_t$ is pre-set as $P_{0}=0$. Vector $\boldsymbol{\beta}=(\beta_0,\beta_1,\ldots,\beta_r) \in \mathbb{R}^{r+1}$ comprises the coefficients of deterministic component $\mu(t;\boldsymbol{\beta})=\beta_0+\beta_1 t/n + \beta_2 (t/n)^2 + \ldots + \beta_r (t/n)^r$, $r \in \mathbb{N} \cup \{0\}$. The supports of $\alpha_A$ and $\alpha_P$ admit the entire real line, but in practice it may be reasonable to confine them to some finite intervals, say $\alpha_A \in (\underline{\alpha_{A}}, \overline{\alpha_{A}})$ and $\alpha_P \in (\underline{\alpha_{P}} , \overline{\alpha_{P}})$, with the end points set arbitrarily to reflect one's presumption of a rather restrained magnitude of the innovations' impact on $A_t$ and $P_t$.

Let $\textbf{y}=(y_1,y_2,\ldots,y_n) \in \mathbb{R}^n$ comprise the modelled time series. Bayesian estimation hinges upon the posterior distribution of the parameters $p(\boldsymbol{\theta}|\textbf{y})=\frac{p(\textbf{y},\boldsymbol{\theta})}{p(\textbf{y})} \propto  p(\textbf{y}|\boldsymbol{\theta})p(\boldsymbol{\theta})$, where $p(\textbf{y},\boldsymbol{\theta})$ denotes the Bayesian statistical model that constitutes the product of the likelihood (or, sampling distribution), $p(\textbf{y}|\boldsymbol{\theta})$, and the prior distribution, $p(\boldsymbol{\theta})$, while $p(\textbf{y})=\int p(\textbf{y},\boldsymbol{\theta}) \text{d} \boldsymbol{\theta}$ is the marginal data density (or, marginal likelihood), which may be dispensed with in deriving and calculating the posterior.
The likelihood function forms the product of univariate Normal densities:
\begin{equation}
p(\textbf{y}|\boldsymbol{\theta})=\prod\limits_{t=1}^{n}p(y_{t}|\mathcal{F}_{t-1},\boldsymbol{\theta})=\prod\limits_{t=1}^{n}f_{N}(y_{t}| m_{t} ,1/\omega )= \left(\frac{2 \pi}{\omega}\right)^{-n/2} e^{-\frac{\omega}{2}\sum\limits_{t=1}^{n}\left(y_t-m_t\right)^2},
\label{wiary}
\end{equation}
where $\mathcal{F}_{t-1}$ signifies the past of the process up to time $t-1$, while $m_t$ and $1/\omega$ denote the conditional mean and variance, respectively, with $ m_t (\boldsymbol{\theta}) \equiv E(y_{t}|\mathcal{F}_{t-1},\boldsymbol{\theta})=\mu(t)+ (a+A_{t-1}) \sum_{j=1}^k q_j \sin(\lambda_j (t+p_j+P_{t-1}))$. At a given $\boldsymbol{\theta}$, the likelihood is calculated recursively, where for a given $t\in{\{1, 2, ..., n}\}$, the values $A_{t-1}, P_{t-1}$ and $y_{t}$ allow one to compute $\epsilon_t=y_t-m_t$, which is then required to obtain 'current' $A_t$ and $P_t$ through the second and third equation in (\ref{model_root}), correspondingly. Despite its apparently simple form, the likelihood function given in (\ref{wiary}) poses quite some challenges. Firstly, the likelihood is periodic with respect to each of $p_j$s (under the other parameters fixed), with the period lengths dependent on $\lambda_j$s and $q_j$s. Therefore, under the support of $p_j$ initially confined to $[0,\pi/\lambda_j]$, for each $j=1, 2, ..., k$, the likelihood function (and thus, the posterior density) for some real-world data sets may still feature multiple modes. Secondly, note that (\ref{wiary}) is invariant to relabelling of the frequencies and corresponding $q_j$s and $p_j$s, which would result in label-switching while sampling from the posterior distribution, featuring $k!$ perfectly symmetric modes (under a symmetric prior). The problem is notoriously well-known in  mixture modelling, with various approaches to addressing it (see, e.g., \cite{fruhwirth2006finite}). Both issues are discussed in more detail in Subsection B.1 of the Supplementary Material.

In the prior distribution, we assume independence for almost all the parameters, with a single exception of $\boldsymbol{\phi}$, the coordinates of which are bound by the stability restriction and hence are \textit{a priori} dependent. Therefore:
$$p(\boldsymbol{\theta})=\left(\prod\limits_{j=1}^{p}p(A_{1-j})\right)\left(\prod\limits_{j=2}^{k}p(q_{j})\right)\left(\prod\limits_{j=1}^{k}p(p_{j})p(\lambda_{j})\right)\left(\prod\limits_{j=0}^{r}p(\beta_{j})\right)p(a)p(\alpha_A)p( \alpha_P)p( \omega)p(\boldsymbol{\phi}).$$
Let $N$ and $U$ stand for the Normal and Uniform distributions, respectively. By $G(a,b)$ we denote the Gamma distribution with mean $ab$ and variance $ab^2$. Further, let $B(b,c,\underline{d},\overline{d})$ denote the Beta distribution with parameters $b,c>0$, and defined over the interval $(\underline{d},\overline{d})$, $\underline{d}< \overline{d}$ and $\underline{d}, \overline{d} \in \mathbb{R}$, with its probability density function given as
$f_B(x;b,c,\underline{d},\overline{d})=\left(\overline{d}-\underline{d}\right)^{-b} \left(x-\underline{d}\right)^{b-1}\left(1-\frac{x-\underline{d}}{\overline{d}-\underline{d}}\right)^{c-1}\Big/Beta(b,c)$
on $x \in (\underline{d},\overline{d})$ and zero otherwise, where $Beta(\cdot,\cdot)$ denotes the Beta function. The prior structure is specified as follows:
\begin{itemize}
\item $\lambda_j \sim B(b_{\lambda_j},c_{\lambda_j},\underline{\lambda}_j,\overline{\lambda}_j)$, $j=1,2,\ldots,k$,
\item $A_{1-j} \sim N( \mu_{A}, \sigma_{A}^2)$, $j=1,2,\ldots,p$,
\item $a \sim N( \mu_{a}, \sigma_{a}^2)$, $q_j \sim N( \mu_{q}, \sigma_{q}^2)$, $j=2,3,\ldots,k$,
\item $p_j \sim B(b_{p_j},c_{p_j},\underline{p}_j,\overline{p}_j)$, $j=1,2,\ldots,k$, where $\overline{p}_j-\underline{p}_j\ge\frac{\pi}{\underline{\lambda}_j}$,
\item $\beta_j \sim N( \mu_{\beta}, \sigma_{\beta}^2)$, $j=0,1,2,\ldots,r$,
\item $\alpha_{A} \sim B(b_{\alpha_{A}},c_{\alpha_{A}}, \underline{\alpha_{A}}, \overline{\alpha_{A}})$,
\item $\alpha_{P} \sim B(b_{\alpha_{P}},c_{\alpha_{P}}, \underline{\alpha_{P}}, \overline{\alpha_{P}})$,
\item $\omega \sim G(a_{\omega},b_{\omega})$,
\item $\rho_j \sim B(b_{\rho},c_{\rho}, \underline{\rho}, \overline{\rho}) $, $j=1, 2, ..., p$,
\end{itemize}
where $\rho_1,\rho_2,\ldots, \rho_p$ denote the vector of partial autocorrelation coefficients of $A_t$ into which we transform the original autoregression coefficients $\boldsymbol{\phi}$ ($via$ the algorithm designed by \cite{Monahan_84}).

Due to the model's complex structure, the posterior distribution does not admit any known form, thereby necessitating a numerical simulation approach. To this end, we combine two most popular Markov Chain Monte Carlo (MCMC) methods: the Gibbs sampler (for $\omega$, which is the only parameter, the conditional posterior of which belongs to a well-known distribution family) with the Random Walk Metropolis-Hastings algorithm (for all the other parameters, the conditional posteriors of which do not belong to known distribution families, thus precluding the Gibbs sampler). A detailed presentation of the algorithm is deferred to Subsection B.1 of the Supplementary Material, while in Subsection B.2 therein we deliver a simulated-data illustration validating the MCMC routine.

\section{Analysis of the Polish GDP}

\begin{figure}[H]
\begin{center}
\subfigure[]{\includegraphics[width=4.53195 cm]{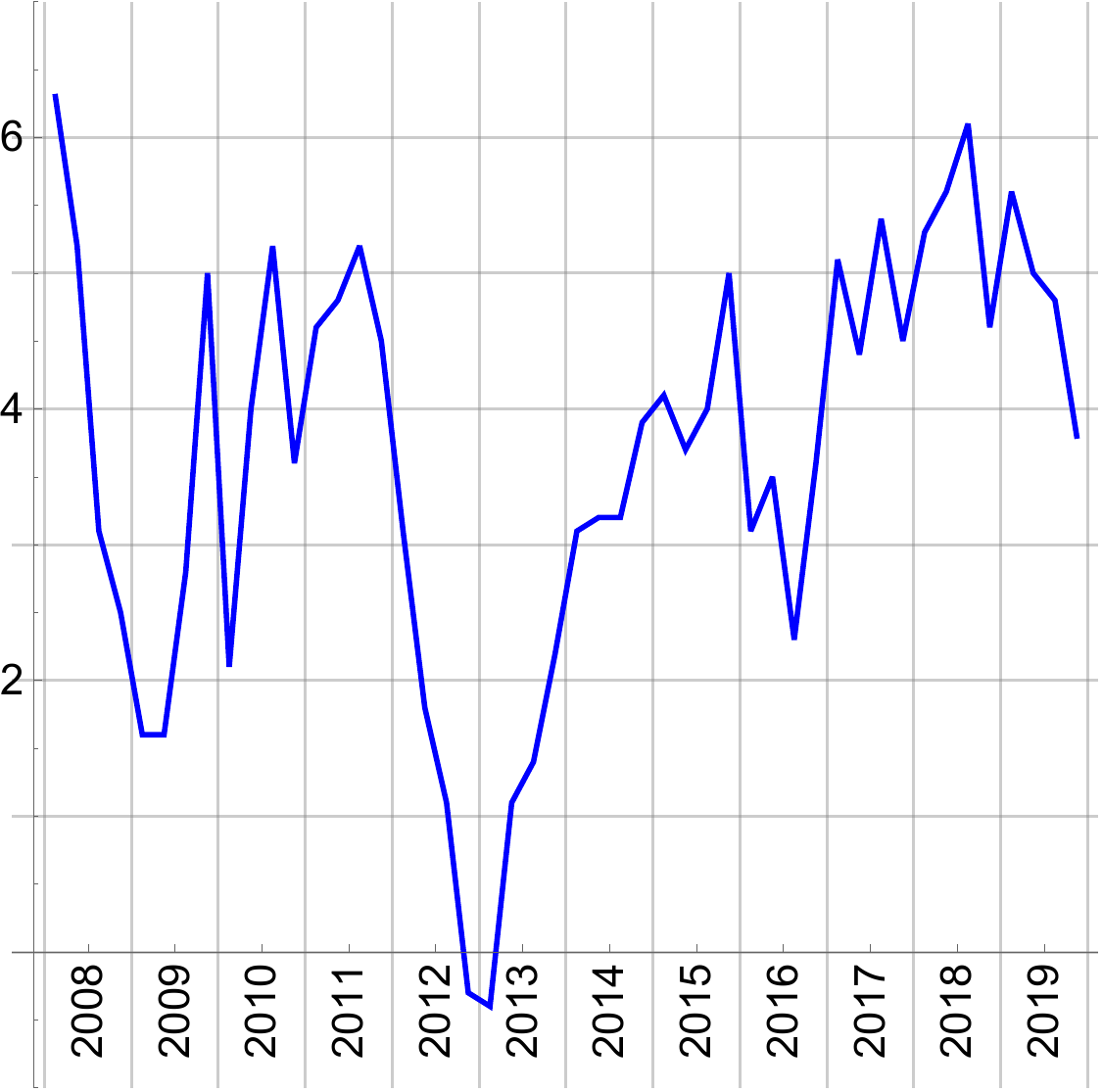}}
\subfigure[]{\includegraphics[width=4.53195 cm]{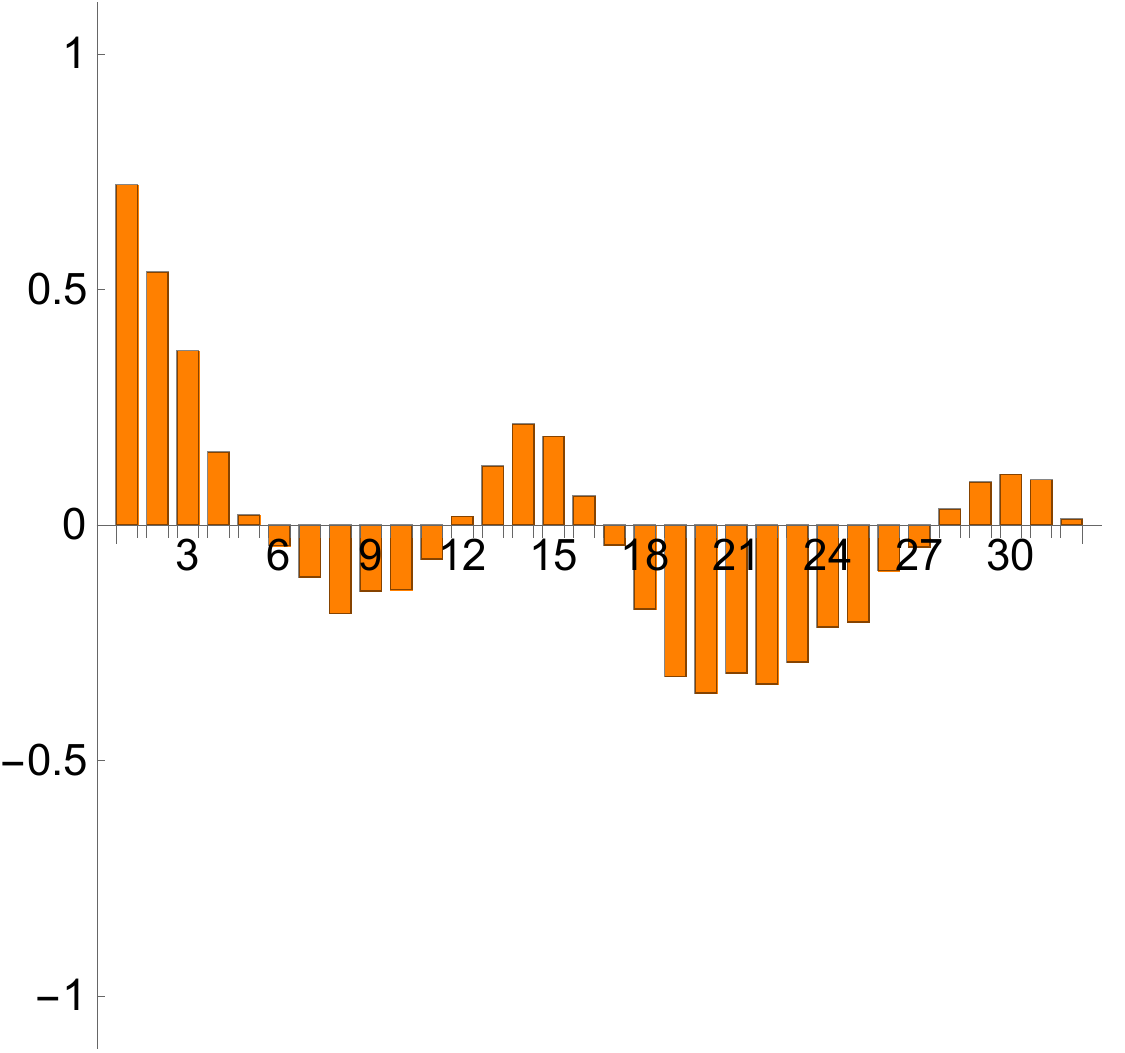}}
\subfigure[]{\includegraphics[width=4.53195 cm]{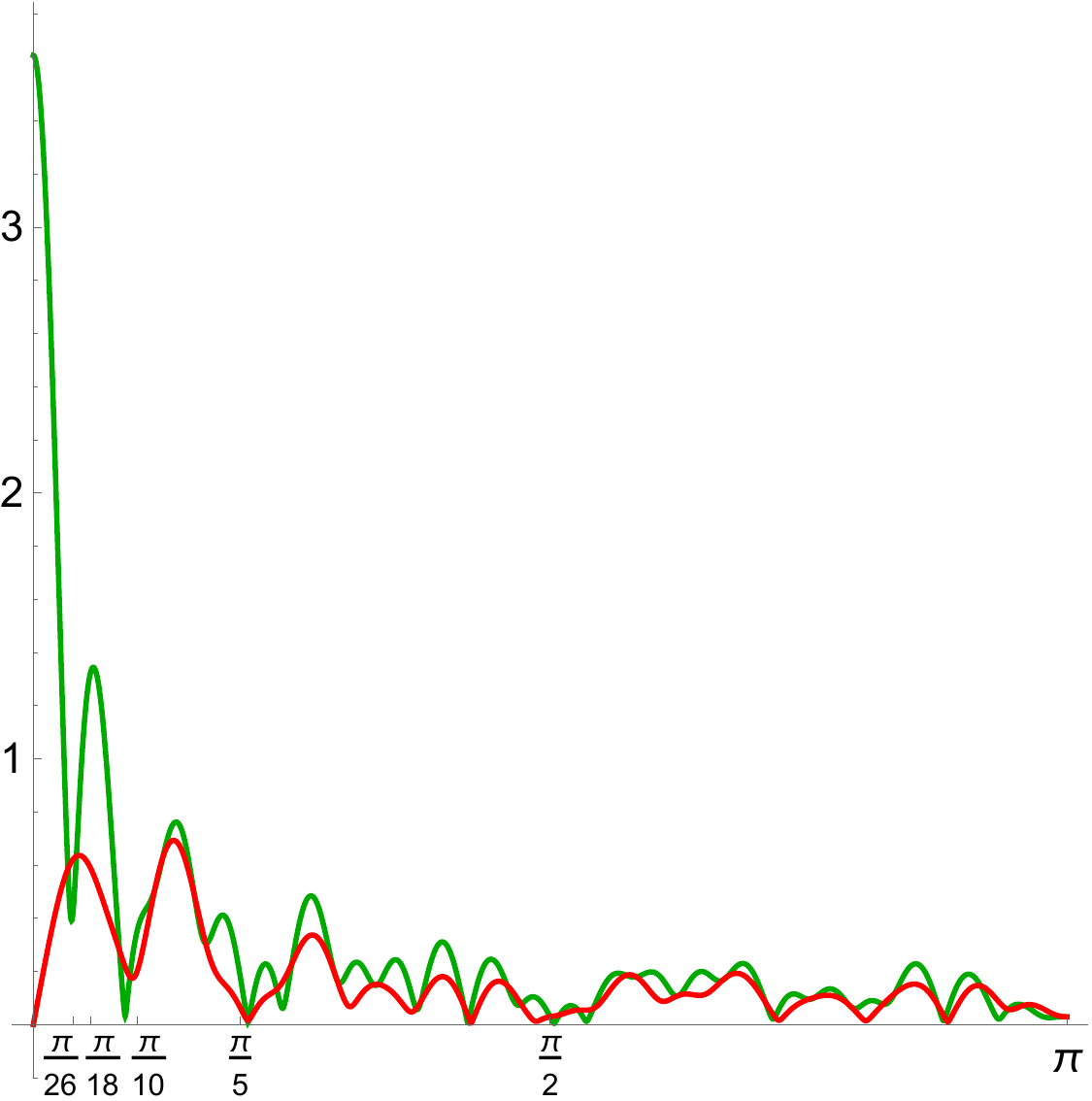}}
\end{center}
\vspace{-0.5 cm}
\caption{Polish quarterly GDP growth rates (y-o-y; panel (\textbf{a})), the sample autocorrelation function (panel (\textbf{b})) and periodograms (panel (\textbf{c})) for the raw data (green line) and demeaned series (red line).}
\label{fig:Real:GDP}
\end{figure}

We use the methodology developed above to analyse the cyclicality of Poland's quarterly GDP year-over-year growth rates at market prices (seasonally and calendar unadjusted data, obtained from Eurostat at \url{https://appsso.eurostat.ec.europa.eu/nui/show.do?dataset=namq\_10\_gdp\&lang=en)}. The series covers the period from Q1.2008 to Q4.2019, which gives the sample size $n=48$. As seen in Figure \ref{fig:Real:GDP}, the data reveal a clear cyclical pattern (reflected also in the ACF; see panel (\textbf{b}) of the figure), with two the most conspicuous troughs around 2008-2009 and 2012-2013, corresponding to the global financial crisis and technical recession in Poland, respectively. The demeaned-data periodogram displays two pronounced peeks at $\hat{\lambda}_1=0.42616$ and $\hat{\lambda}_2=0.13895$, corresponding to the cycle lengths of $T(\hat{\lambda}_1)=2 \pi/(4 \hat{\lambda}_1)=3.686$ years (the Kitchin inventory cycle) and $T(\hat{\lambda}_2)=2 \pi/(4 \hat{\lambda}_2)=11.305$ years (the Juglar investment cycle). Note that, similarly as in the simulation data study (presented in Subsection B.2 of the Supplementary Material, we associate the frequencies with the periodogram maxima in the descending order of the periodogram values, thus $\lambda_1>\lambda_2$. 

\begin{figure}[H]
\begin{center}
\subfigure[Amplitude $A_t+a$]{\includegraphics[width=7 cm, height=2.234505 cm]{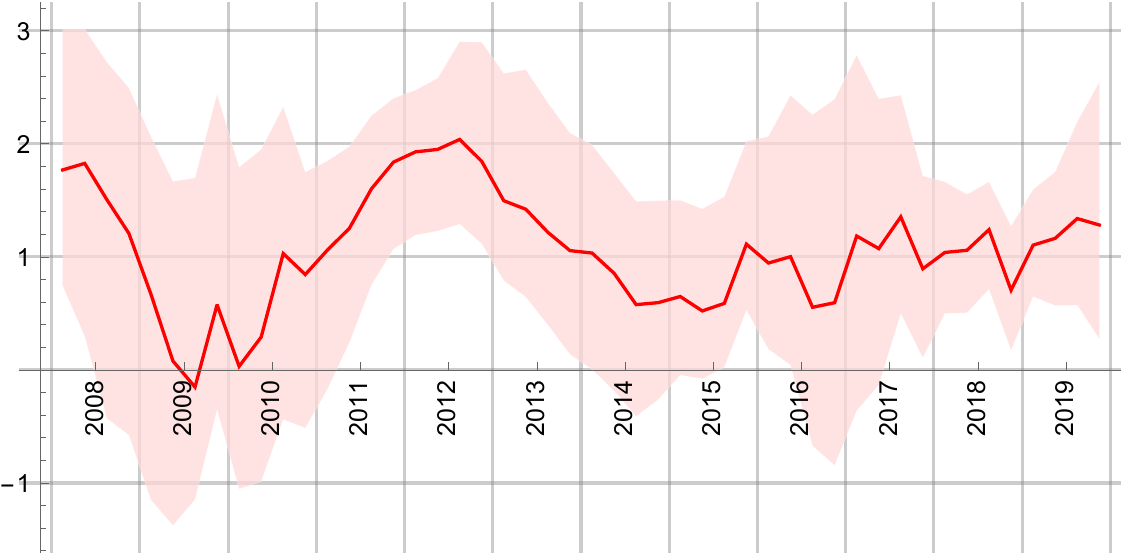}}\hspace{0.2 cm}
\subfigure[Phase shifts $P_t$]{\includegraphics[width=7 cm, height=2.234505 cm]{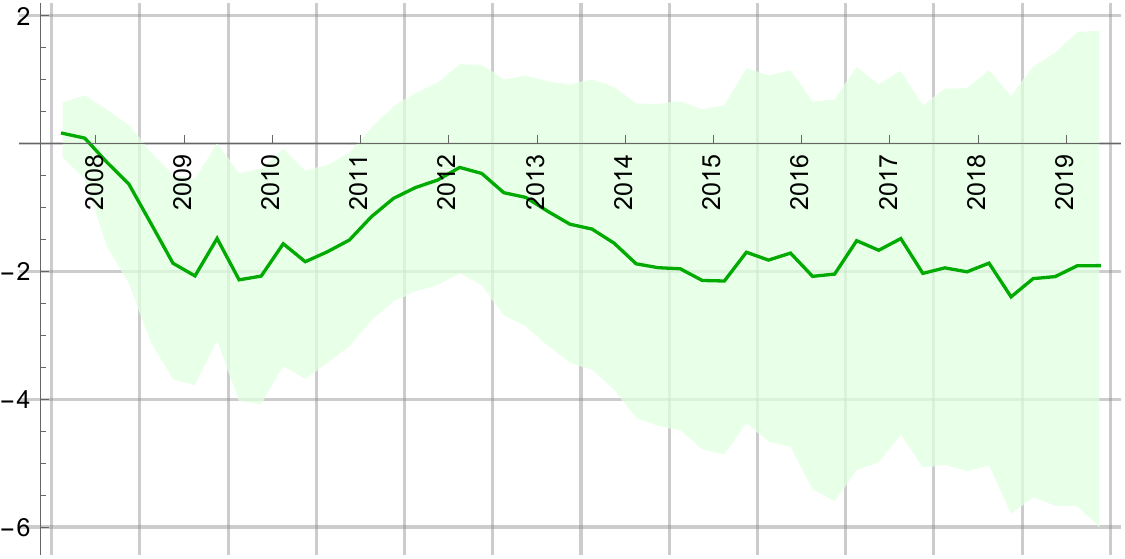}}
\end{center}
\vspace{-0.15 cm}
\caption{Posterior medians of the amplitude (a) and phase shift (b), with the 95\% credible intervals.}
\label{fig:Real:amplitude_phase_shifts}
\end{figure}

The MCMC starting points for the frequencies, $\hat{\lambda}_1=0.42616$ and $\hat{\lambda}_2=0.13895$, retrieved from the demeaned-data periodogram, are relatively far from the posterior location measures: $E(\lambda_1|y)\approx Me(\lambda_1|y) \approx Mo(\lambda_1|y)\approx	0.46$ and $E(\lambda_2|y)\approx Me(\lambda_2|y) \approx Mo(\lambda_2|y)\approx	0.1477$, respectively (see Table \ref{Table:Real:poster_chars}).Details on the settings for the prior distribution, the MCMC algorithm and its convergence are provided in Section D.1 of the Supplementary Material. The posterior histograms of individual parameters and additionally, bivariate posteriors are presented in Section D.2 therein. Specifically, $\hat{\lambda}_1$ is nearing the left ends of the 95\% credible and HPD intervals, while $\hat{\lambda}_2$ actually falls beyond these (see Table \ref{Table:Real:poster_chars}). Low posterior uncertainty of the frequencies translates to also low dispersion of the posteriors induced for the cycles' lengths (in years), $T(\lambda_j)=2\pi/(4\lambda_j)$, $j=1, 2$; see Table \ref{Table:Real:poster_chars} and also Figure~D.9 in the Supplementary Material. Based on the HPD intervals, we can infer that, with 0.95 posterior probability, the first cycle lasts between 3.15 and 3.69 years, whereas the second one - between 10.56 and 10.71 years. The results agree neatly with the Kitchin inventory cycle and the Juglar investment cycle. However, as regards the relative 'weights' of the both, it cannot be clearly determined which of the two business fluctuation types actually dominates. The posterior density of $q_2$ features two modes (both covered by the 95\% HPD interval) implying contradictory results (see Table \ref{Table:Real:poster_chars}). The dominant one, $Mo_1(q_2|y)\approx -0.6617$, indicates that it is the Kitchin cycle that prevails in the modelled series, with the relative weight of the Juglar fluctuations accounting for around $|Mo_1(q_2|y)|\approx 0.6617$ of the inventory cycle. However, the value of the other, lower mode, $Mo_2(q_2|y)\approx -1.1488 $, would suggest somewhat the opposite, that is that the investment cycle is the leading one, constituting $|Mo_2(q_2|y)|\approx 1.1488$ of the inventory fluctuations. Due to the above, and combined with a generally rather high posterior uncertainty of the parameter, no clear-cut inference about the relative dominance of the identified cycles is available here. Even more so due to fairly close posterior chances of $q_2 \in (-1, 0)$ and $q_2 <-1$, equal to 0.565 and 0.435, respectively.
Figure \ref{fig:Real:amplitude_phase_shifts} presents the posterior medians of the amplitude ($A_t+a$) and phase ($P_t$) shifts, along with the bands of 95\% credible intervals. We notice a strong positive correlation between the two series of posterior medians, which is an artefact of the model's construction (both processes driven by the same error term, $\epsilon_t$), as well as high persistence of the amplitude shifts ($Mo(\phi_1|y)\approx 0.96$; see Table \ref{Table:Real:poster_chars}). Additionally, both processes share a very similar posterior distribution of their corresponding noise scaling parameters, that is $\alpha_A$ and $\alpha_P$ (\textcolor{black}{see Figure D.7 in the Supplementary Material}). The uncertainty related to the phase shifts tends to raise gradually over time, which can be ascribed to the $P_t$ process specified as a random walk. The results bear a close resemblance to the ones obtained in the simulated-data example presented in Subsection B.2 of the Supplementary Material.

\begin{figure}[H]
\begin{center}
\subfigure[Juglar investment cyclic component]{\includegraphics[width=7 cm, height=2.234505 cm]{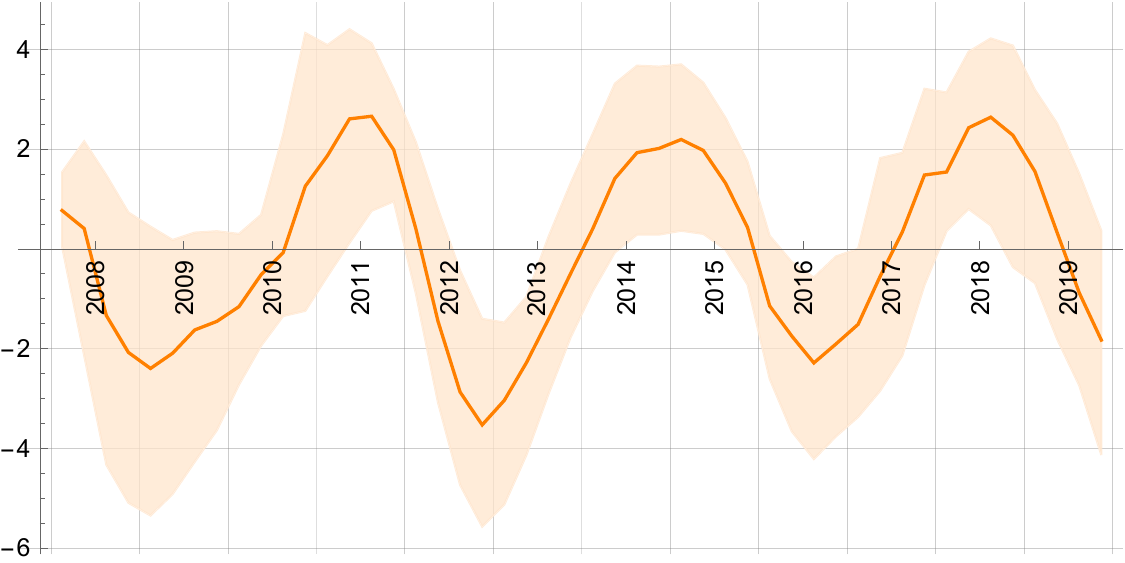}}\hspace{0.2 cm}
\subfigure[Kitchin inventory cyclic component]{\includegraphics[width=7 cm, height=2.234505 cm]{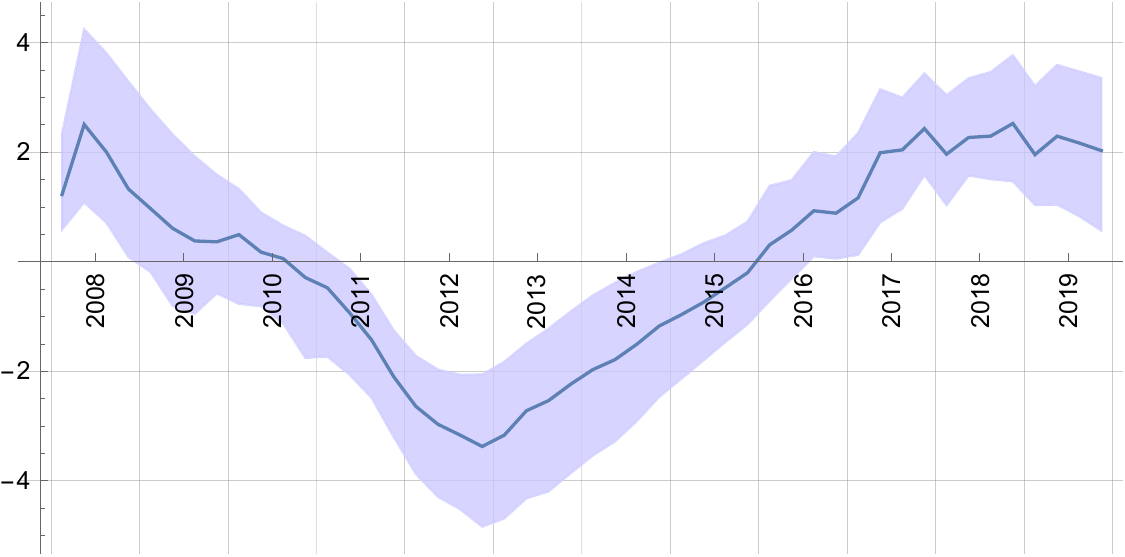}}
\end{center}
\vspace{-0.275 cm}
\caption{Posterior medians, with 95\% credible intervals, of the Kitchin and Juglar components.}
\label{fig:Real:cycle1_2}
\end{figure}

Time-varying amplitude and length of business fluctuations for certain cyclical indicators are recognized as ones of typical characteristics of business cycles (see \cite{Zarnowitz_1992}). Usually, one of the problems encountered in business cycle modelling is not the lack of sufficient sample size, but accounting for this variability by allowing for it explicitly in the model's specification, as it is done in this paper. Here, the overall cyclicality of a modelled series is specified as a sum of $k$ components of the form $C_{j,t}= (a+A_{t-1})q_j\sin(\lambda_j (t+p_j+P_{t-1}))$, corresponding to frequencies $\lambda_j$ for $j=1,...,k$, with $k$ being pre-determined on the basis of demeaned-data periodogram. Under the empirical results obtained above, the Kitchin cycle component, related to $\lambda_1$, is governed by the formula ($q_1\equiv 1$): $K_t \equiv C_{1,t} = (a+A_{t-1}) \sin(\lambda_1 (t+p_1+P_{t-1}))$, while the Juglar component, related to $\lambda_2$, by: $J_t \equiv C_{2,t} = (a+A_{t-1}) q_2 \sin(\lambda_2 (t+p_2+P_{t-1}))$, so the observable process decomposes as $y_t=K_t+J_t+\mu(t)+\epsilon_t$. Thus, inference can be conducted about each of the cycles individually. The Bayesian estimation approach enables us to infer not only about the 'central' trajectories of each component (through relevant posterior location measures of $K_t$ and $J_t$), but also, and more importantly, about the uncertainty at each time $t=1, \dots, n$. In Figure \ref{fig:Real:cycle1_2} we present the posterior medians of both the cycles, $Me(K_t|y)$ and $Me(J_t|y)$, along with the band of $95\%$ quantile credible intervals. The posterior median trajectory of the inventory cycle reveals smooth oscillations featuring time-varying (but not rapidly) amplitude and phase shifts, the changes of which neatly correspond with the dynamics of $A_t+a$ and $P_t$, respectively, as seen in Figure \ref{fig:Real:amplitude_phase_shifts}. The credible intervals along the Kitchin component's posterior medians display time-varying uncertainty, with a clear and intuitive pattern of its increases at the turning points (peaks and troughs), alternating with decreases along the periods of same-direction consecutive movements. Similar observations could be made with respect to the Juglar fluctuations, but only to a much limited extent (see Figure \ref{fig:Real:cycle1_2} (b)), which can to attributed to the sample range covering only one full-length Juglar cycle run. Therefore, it can be concluded that the dynamics of the amplitude and phase shift changes, presented in Figure \ref{fig:Real:amplitude_phase_shifts}, pertains predominantly to the shorter, inventory (rather then the longer, investment) fluctuations. Nevertheless, both of the identified cycles reveal clear asymmetry of the contraction and expansion phases, with the upswings lasting longer than the preceding downslides, which remains in accord with the literature (see, e.g., \cite{Neftci_1984}, \cite{Falk_1986}, \cite{Sichel_1993}, \cite{krolzig2013markov}, \cite{Morley_2012}, \cite{Chini_2018}).  

\begin{figure}[H]
\includegraphics[width=3.2071 cm, height=2.05 cm]{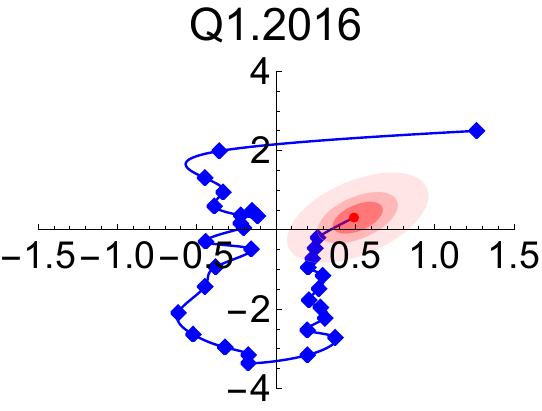}
\includegraphics[width=3.2071 cm, height=2.05 cm]{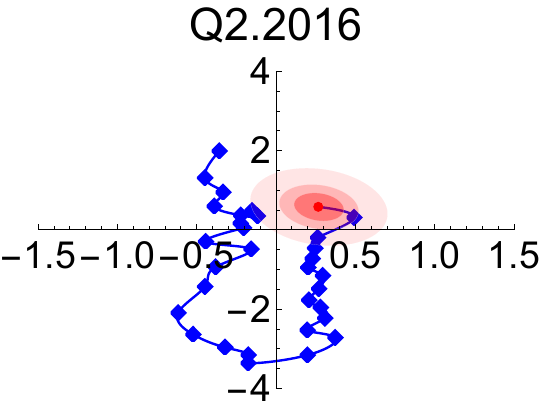}
\includegraphics[width=3.2071 cm, height=2.05 cm]{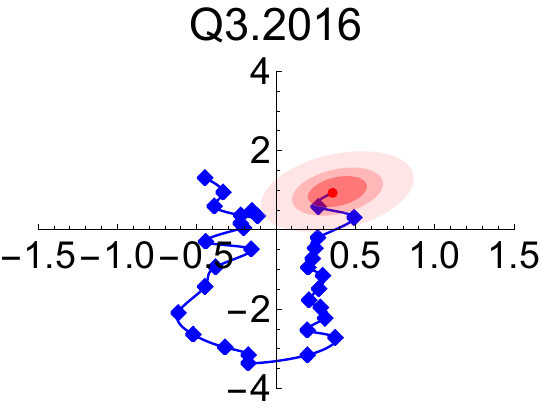}
\includegraphics[width=3.2071 cm, height=2.05 cm]{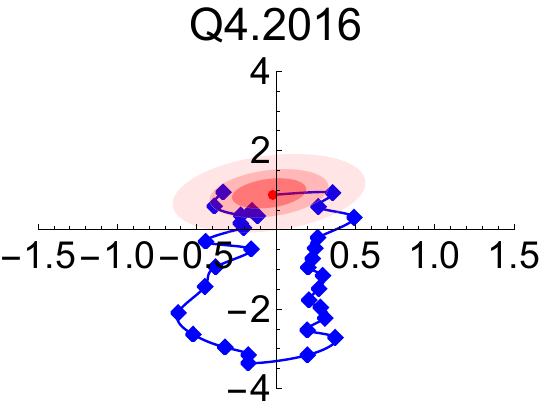}
\includegraphics[width=3.2071 cm, height=2.05 cm]{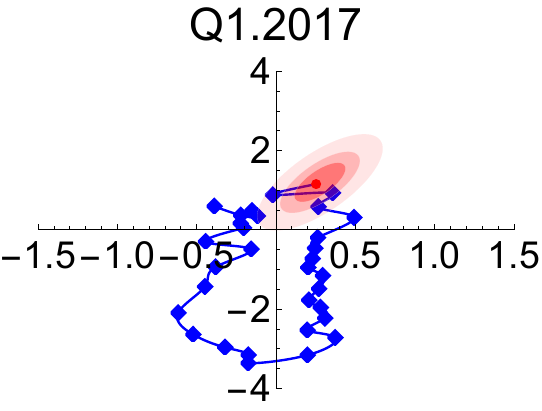}
\includegraphics[width=3.2071 cm, height=2.05 cm]{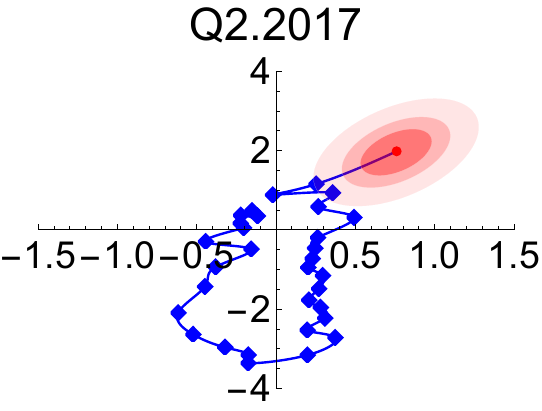}
\includegraphics[width=3.2071 cm, height=2.05 cm]{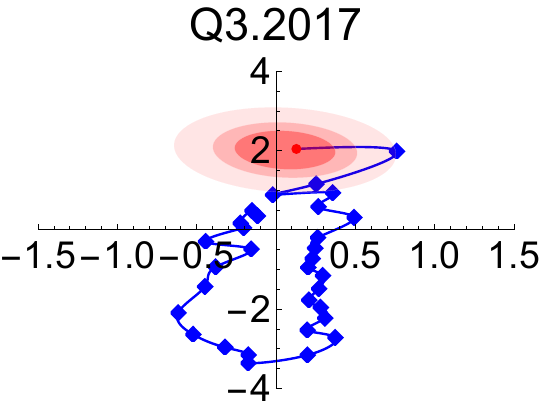}
\includegraphics[width=3.2071 cm, height=2.05 cm]{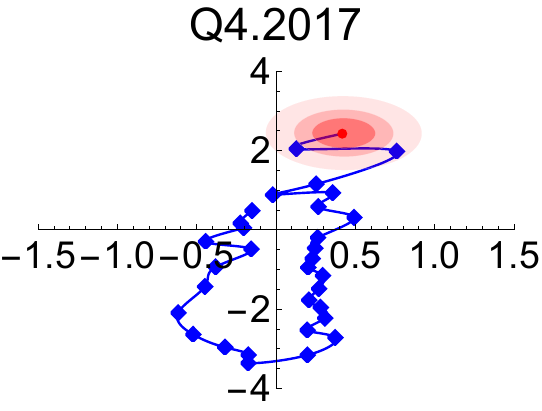}
\includegraphics[width=3.2071 cm, height=2.05 cm]{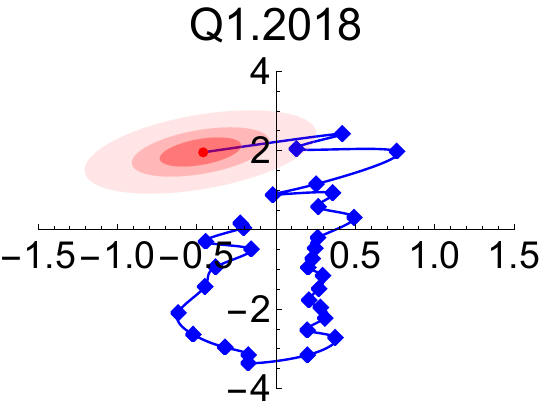}
\includegraphics[width=3.2071 cm, height=2.05 cm]{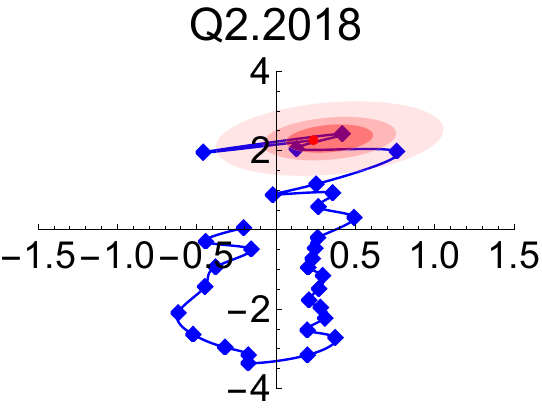}
\includegraphics[width=3.2071 cm, height=2.05 cm]{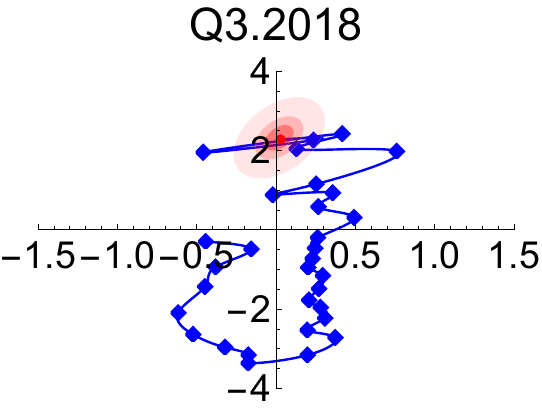}
\includegraphics[width=3.2071 cm, height=2.05 cm]{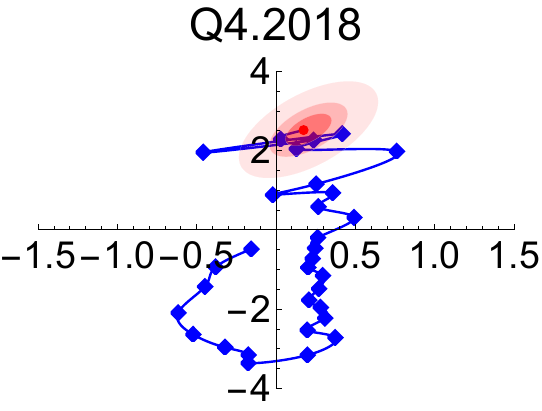}
\includegraphics[width=3.2071 cm, height=2.05 cm]{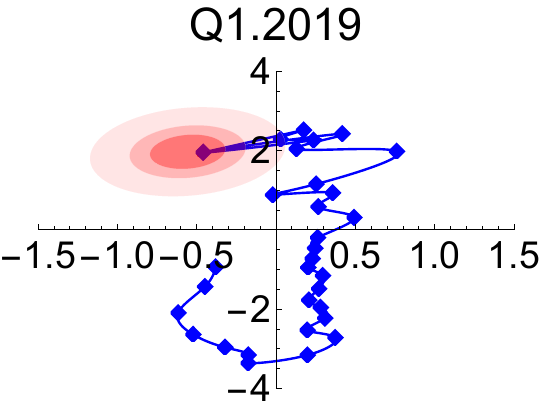}\hspace{0.9765 cm}
\includegraphics[width=3.2071 cm, height=2.05 cm]{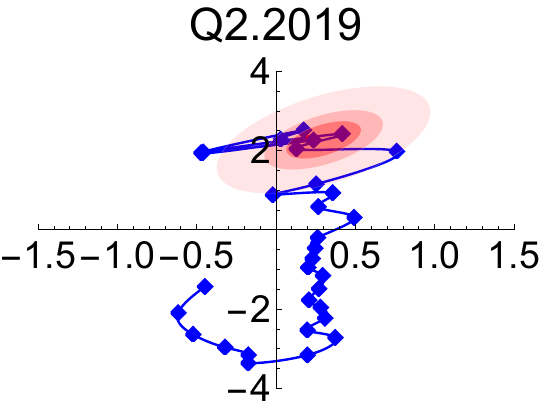}\hspace{0.9765 cm}
\includegraphics[width=3.2071 cm, height=2.05 cm]{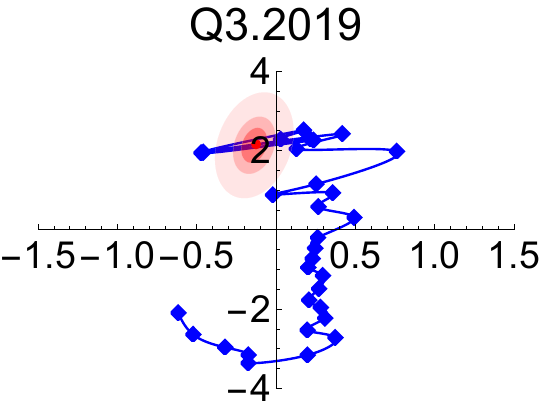}\hspace{0.9765 cm}
\includegraphics[width=3.2071 cm, height=2.05 cm]{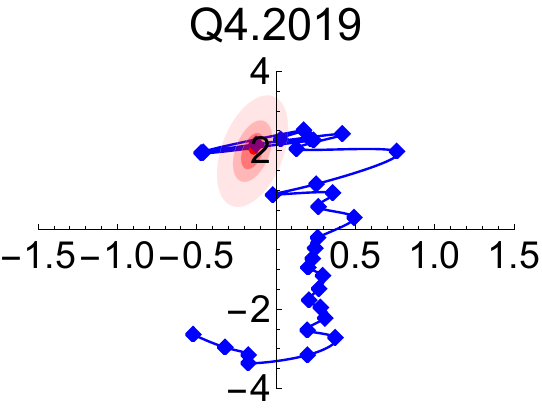}\\
\vspace{-0.5 cm}
\caption{Evolution over time of the posterior medians of the Juglar business cycle clocks (blue lines), displayed for 8-year rolling windows, with 30, 60 and 90\% quantile ellipsoids at the final data point, starting from Q1.2016 to Q4.2019 (date given in each panel's heading).}
\label{fig:Real:clocks2}
\end{figure}

\begin{figure}[H]
\includegraphics[width=3.2071 cm, height=2.05 cm]{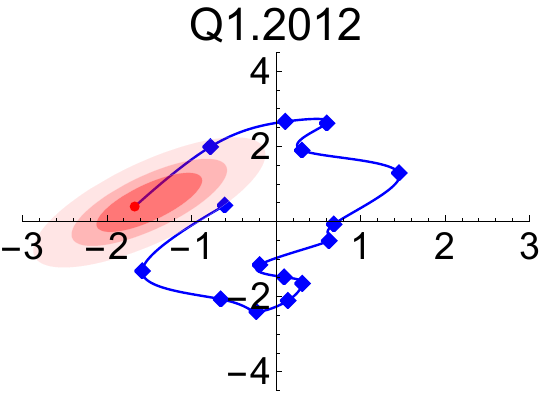}
\includegraphics[width=3.2071 cm, height=2.05 cm]{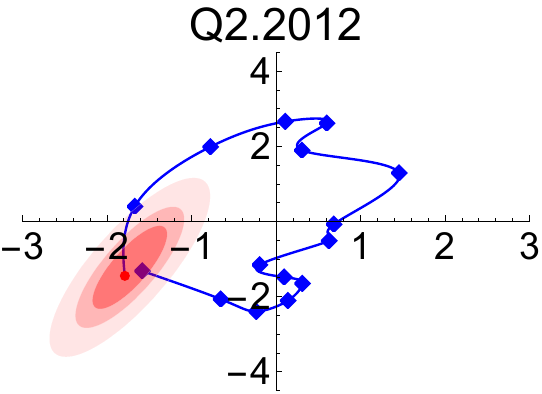}
\includegraphics[width=3.2071 cm, height=2.05 cm]{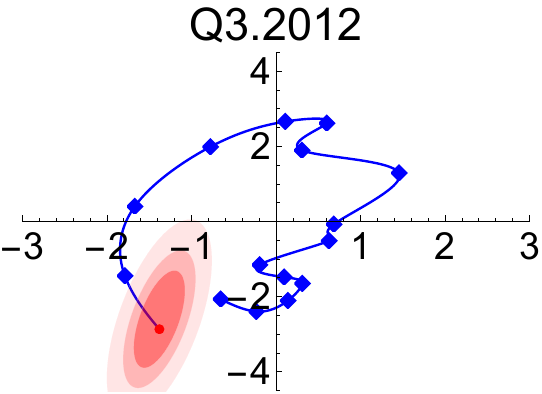}
\includegraphics[width=3.2071 cm, height=2.05 cm]{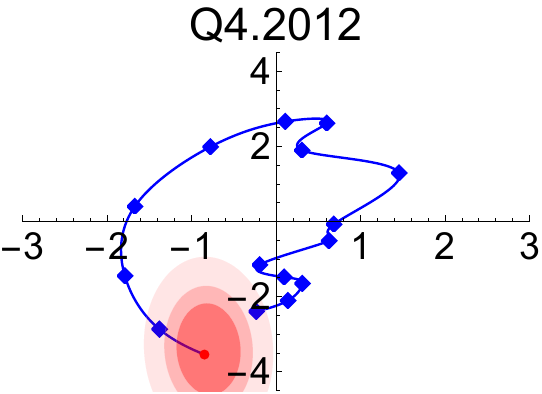}
\includegraphics[width=3.2071 cm, height=2.05 cm]{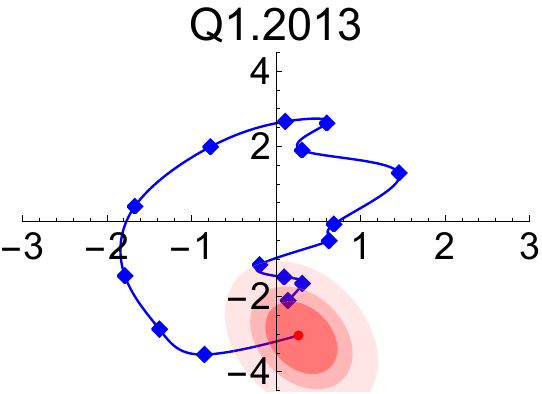}
\includegraphics[width=3.2071 cm, height=2.05 cm]{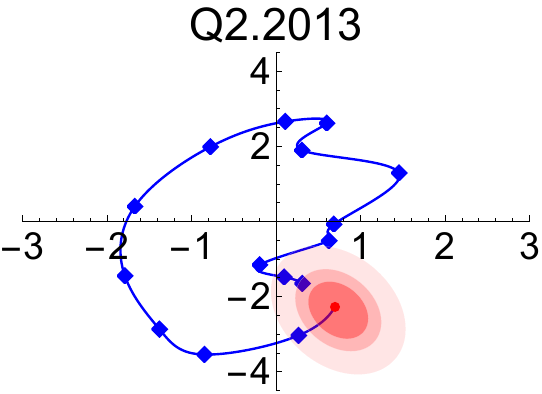}
\includegraphics[width=3.2071 cm, height=2.05 cm]{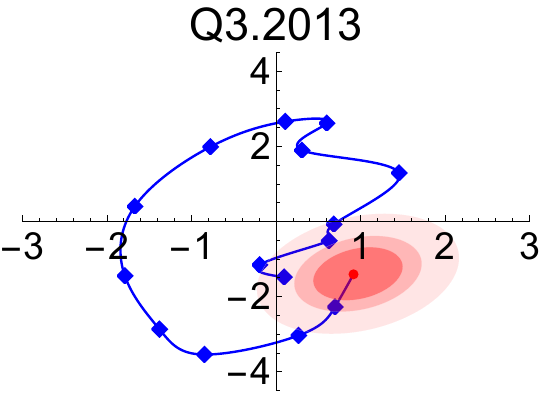}
\includegraphics[width=3.2071 cm, height=2.05 cm]{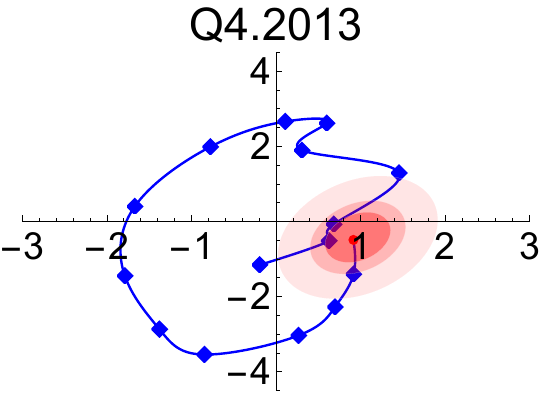}
\includegraphics[width=3.2071 cm, height=2.05 cm]{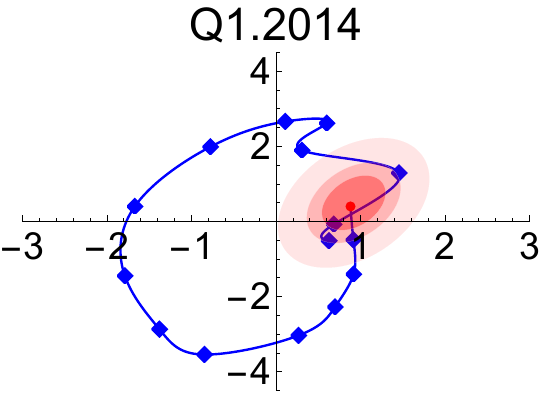}
\includegraphics[width=3.2071 cm, height=2.05 cm]{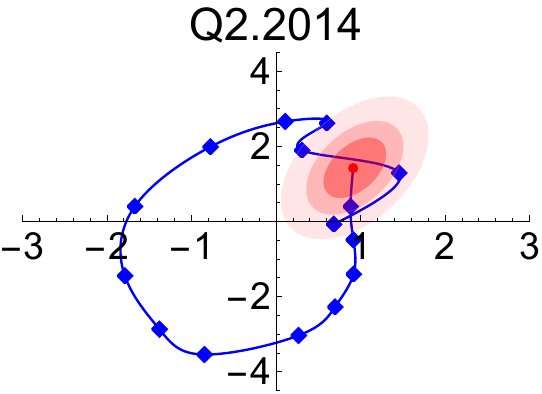}
\includegraphics[width=3.2071 cm, height=2.05 cm]{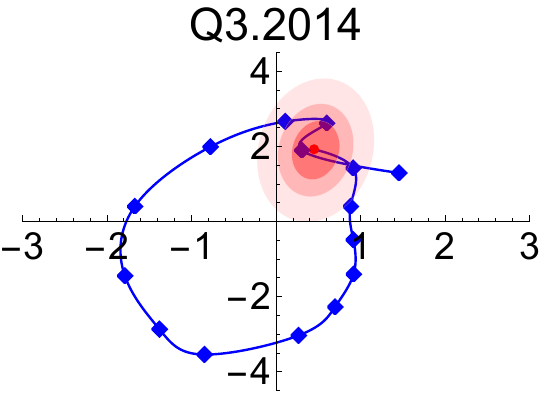}
\includegraphics[width=3.2071 cm, height=2.05 cm]{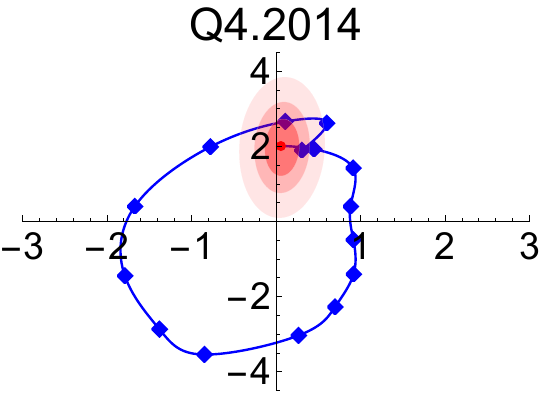}
\includegraphics[width=3.2071 cm, height=2.05 cm]{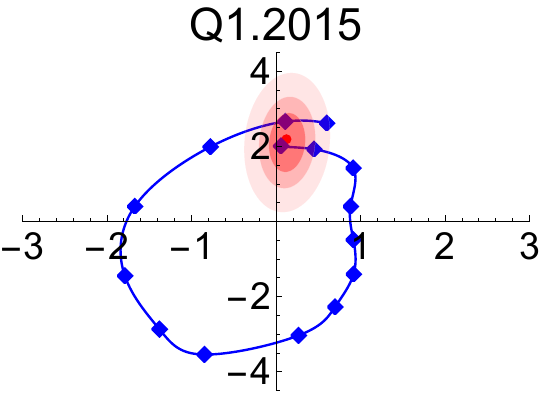}
\includegraphics[width=3.2071 cm, height=2.05 cm]{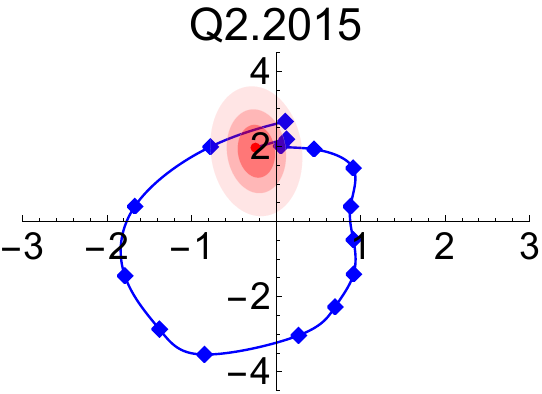}
\includegraphics[width=3.2071 cm, height=2.05 cm]{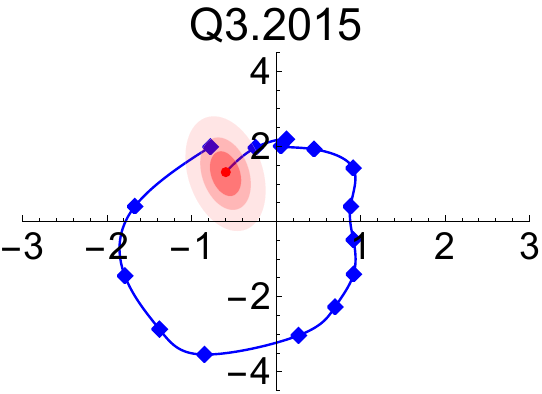}
\includegraphics[width=3.2071 cm, height=2.05 cm]{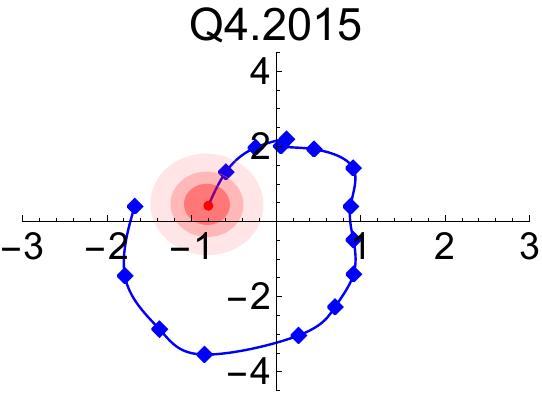}
\includegraphics[width=3.2071 cm, height=2.05 cm]{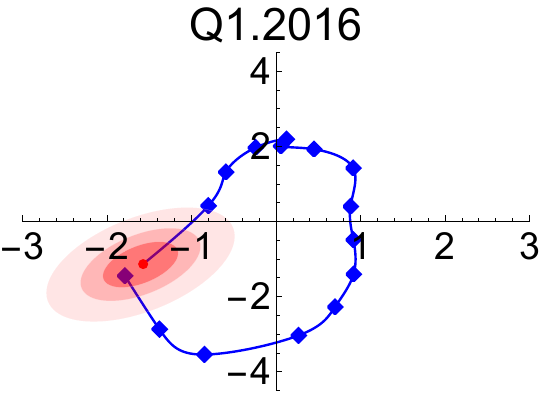}
\includegraphics[width=3.2071 cm, height=2.05 cm]{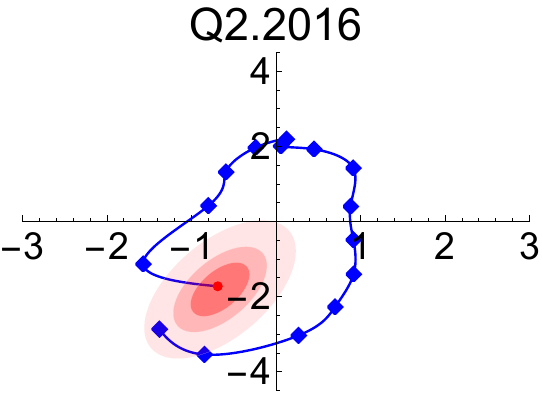}
\includegraphics[width=3.2071 cm, height=2.05 cm]{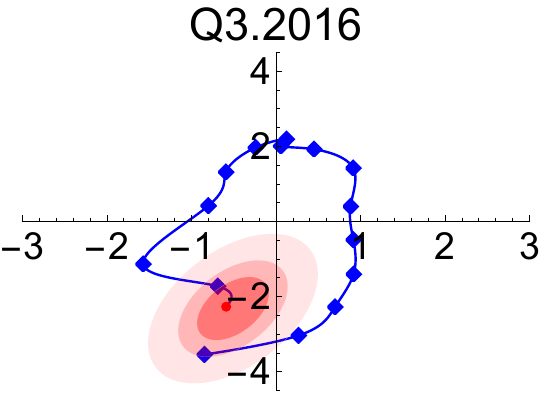}
\includegraphics[width=3.2071 cm, height=2.05 cm]{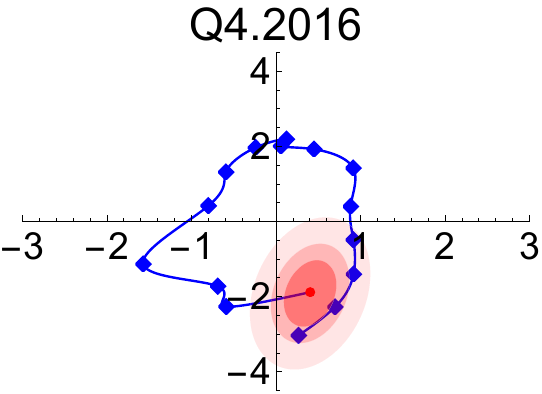}
\includegraphics[width=3.2071 cm, height=2.05 cm]{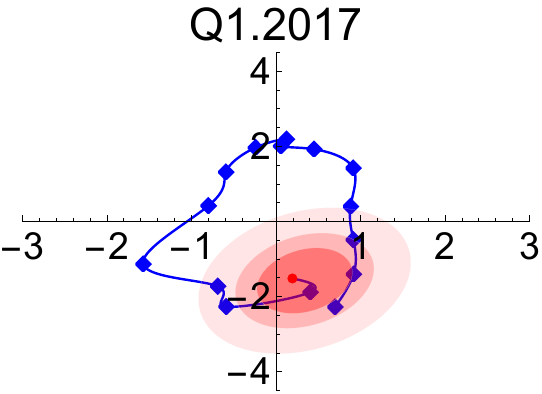}
\includegraphics[width=3.2071 cm, height=2.05 cm]{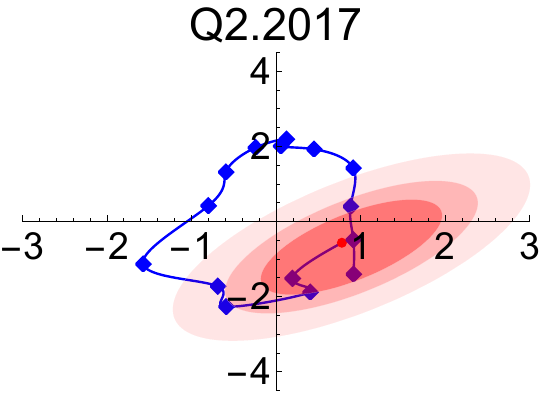}
\includegraphics[width=3.2071 cm, height=2.05 cm]{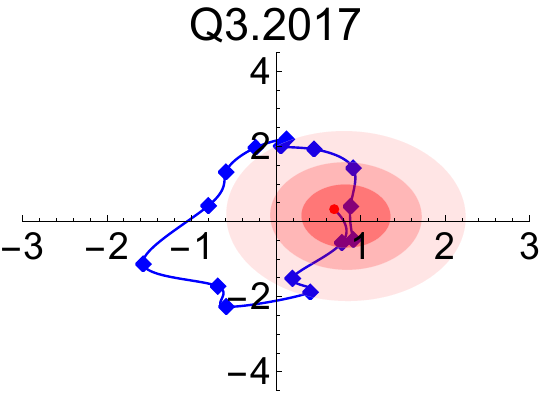}
\includegraphics[width=3.2071 cm, height=2.05 cm]{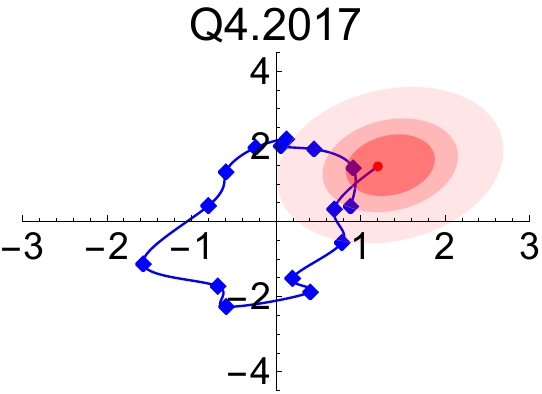}
\includegraphics[width=3.2071 cm, height=2.05 cm]{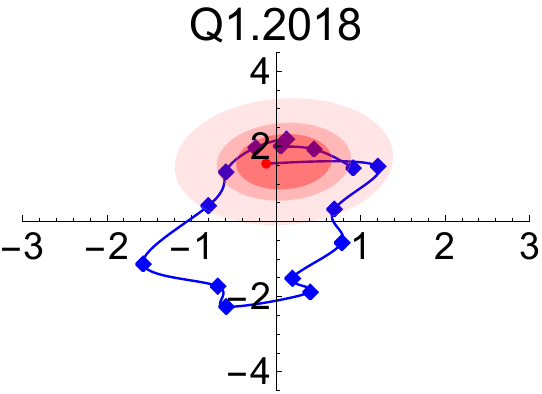}
\includegraphics[width=3.2071 cm, height=2.05 cm]{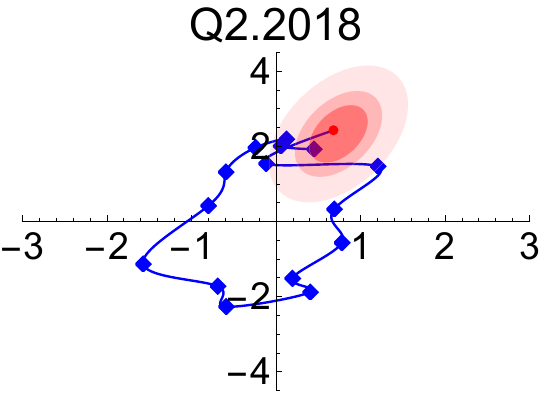}
\includegraphics[width=3.2071 cm, height=2.05 cm]{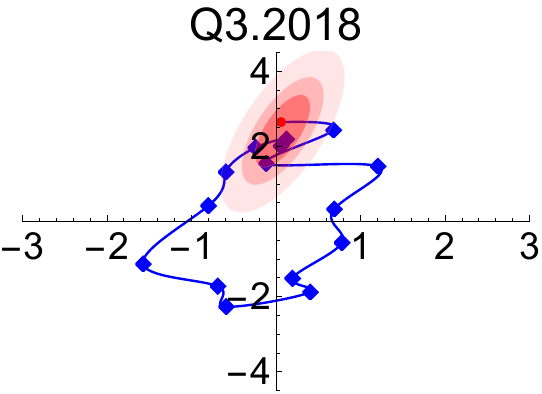}
\includegraphics[width=3.2071 cm, height=2.05 cm]{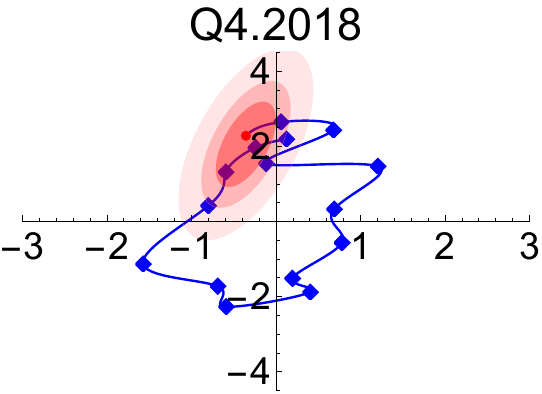}
\includegraphics[width=3.2071 cm, height=2.05 cm]{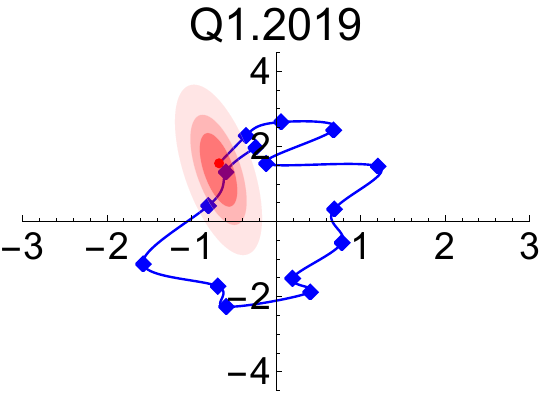}\hspace{0.9765 cm}
\includegraphics[width=3.2071 cm, height=2.05 cm]{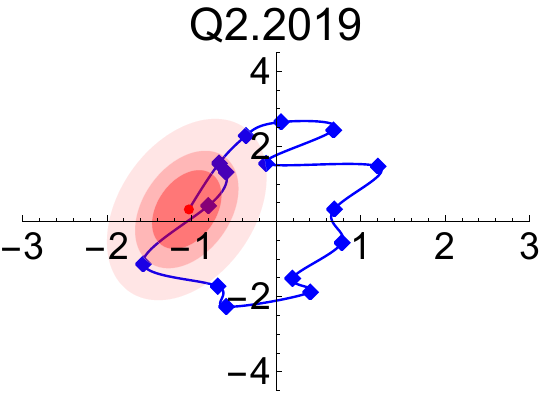}\hspace{0.9765 cm}
\includegraphics[width=3.2071 cm, height=2.05 cm]{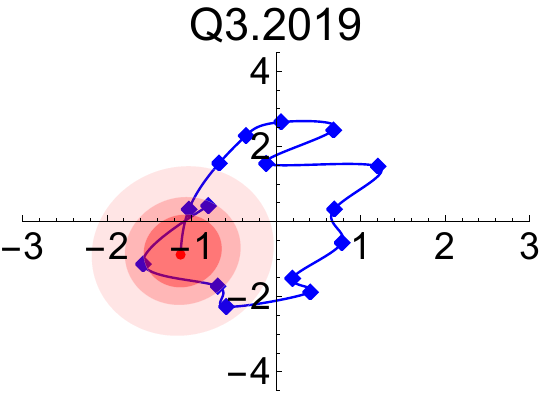}\hspace{0.9765 cm}
\includegraphics[width=3.2071 cm, height=2.05 cm]{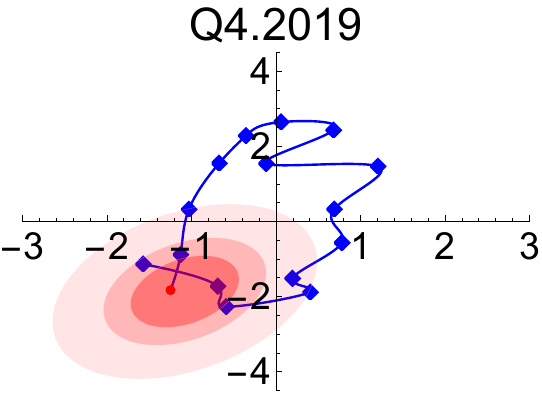}\\
\vspace{-0.05 cm}
\caption{Evolution over time of the posterior medians of the Kitchin business cycle clocks (blue lines), displayed for 4-year rolling windows, with 30, 60 and 90\% quantile ellipsoids at the final data point, over the period from Q1.2012 to Q4.2019 (date given in each panel's heading).}
\label{fig:Real:clocks1}
\end{figure}

\begin{table}[H]
\spacingset{1.532}
\begin{center}
{\footnotesize{
\begin{tabular}{c|cccccc}
Parameter  & Mean & Median & Mode & St. dev. & $\text{CI}_{0.95}$ & $\text{HPD}_{0.95}$ \\\hline
$a$ & 1.568 & 1.550 & 1.660 & 0.778 & (0.411; 3.198) & (0.305; 3.021) \\
$\lambda _1$ & 0.460 & 0.459 & 0.458 & 0.019 & (0.425; 0.498) & (0.424; 0.498) \\
$\lambda _2$ & 0.148 & 0.148 & 0.148 & 0.001 & (0.147; 0.149) & (0.147; 0.149) \\
$\alpha _A$ & 0.411 & 0.413 & 0.426 & 0.108 & (0.188; 0.616) & (0.195; 0.622) \\
$\alpha _P$ & 0.420 & 0.421 & 0.425 & 0.121 & (0.185; 0.651) & (0.188; 0.654) \\
$\phi_1$ & 0.851 & 0.888 & 0.959 & 0.170 & (0.482; 0.987) & (0.629;  0.999) \\
$\beta_0$  & 3.859 & 3.865 & 3.875 & 0.142 & (3.563; 4.125) & (3.571; 4.131) \\
$\omega$  & 1.506 & 1.477 & 1.429 & 0.344 & (0.916; 2.258) & (0.874; 2.198) \\
$A_0$ & 0.0 & 0.0 & 0.0 & 0.001 & (-0.002; 0.002) & (-0.002; 0.002) \\
$p_1$ & 3.957 & 4.546 & 4.942; 2.151 & 1.515 & (0.758; 5.932) & (0.901; 6.038) \\
$p_2$ & -9.159 & -9.312 & -9.464 & 1.776 & (-12.293; -4.973) & (-12.614; -5.362) \\
$q_2$ & -1.097 & -0.898 & -0.662; -1.149 & 0.531 & (-2.325; -0.505) & (-2.098; -0.443) \\
$T(\lambda_1)$ & 3.421 & 3.420 & 3.421 & 0.142 & (3.153; 3.697) & (3.149; 3.691) \\
$T(\lambda_2)$ & 10.637 & 10.636 & 10.637 & 0.039 & (10.560; 10.715) & (10.560; 10.714)\\
\end{tabular}
 \caption{Posterior characteristics of the model's parameters and cycle lengths.}
\label{Table:Real:poster_chars}
}}
\end{center}
\end{table}
Distilling each cycle's individual component from the modelled data enables us to build growth rate cycle clocks for each of them, respectively. Figures \ref{fig:Real:clocks2} and \ref{fig:Real:clocks1} present the series of clocks obtained for the Juglar and Kitchin cycles as the trajectories (blue lines) of points $\left( Me(\Delta J_t |y), Me(J_t |y)\right )$ and $\left( Me(\Delta K_t |y), Me(K_t |y)\right )$, respectively. Moreover, owing to the application of the Bayesian methodology, entire posterior distributions of bivariate random variables $(\Delta J_t, J_t)$ and $(\Delta K_t , K_t)$ can be easily induced, allowing presentation of the uncertainty inherent to each point of the clock. Owing to that, one can formally calculate the posterior probability of each point actually falling into a given quadrant, which enables a fully probabilistic approach to growth rate cycle chronology.

\begin{figure}[H]
 \begin{center}
\subfigure[Kitchin cycle]{\includegraphics[width=6.995 cm]{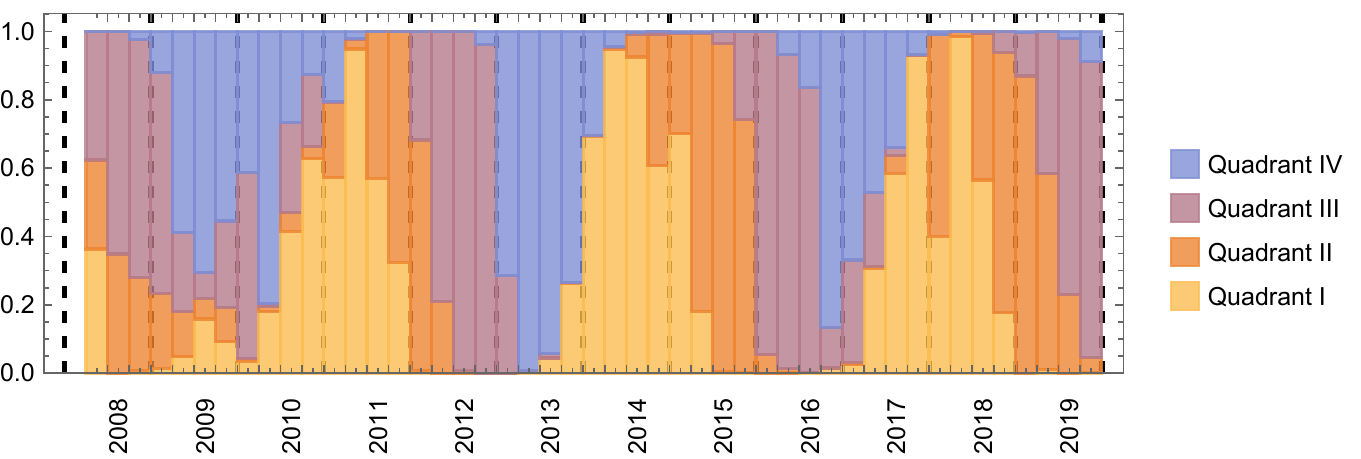}}
\subfigure[Juglar cycle]{\includegraphics[width=6.995 cm]{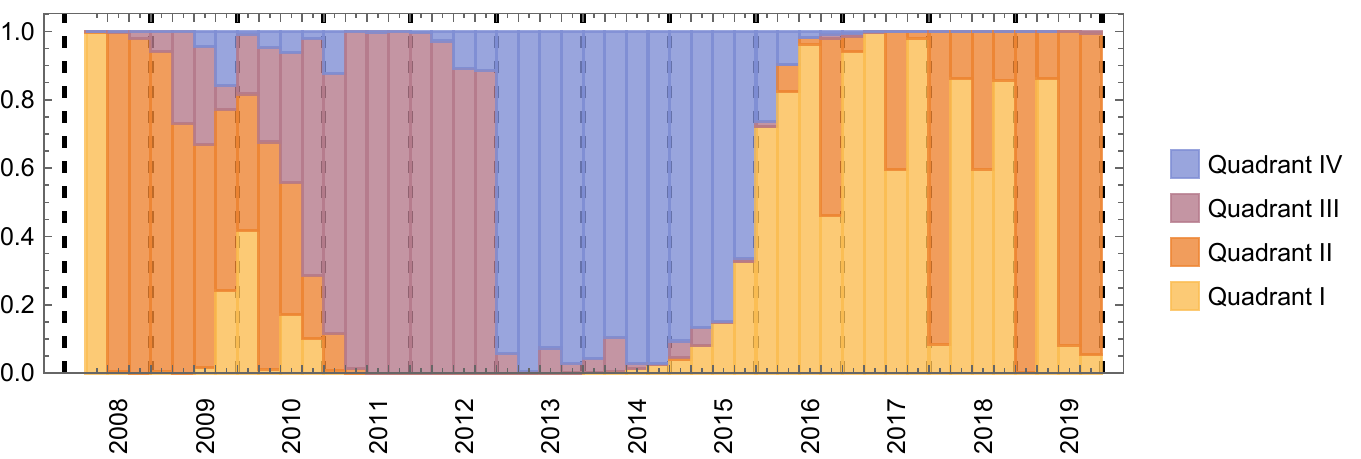}}
\caption{Posterior probabilities of the Kitchin and Juglar cycles clocks' quadrants at each data point.}
\label{fig:Real:cyclesProbs}
\end{center}
\end{figure}

The uncertainty calculated for the cycle clocks' points, based on the posterior MCMC sample induced for pairs $(\Delta J_t, J_t)$ and  $(\Delta K_t , K_t)$ can be visualized by bivariate quantile ellipsoids, as in Figures \ref{fig:Real:clocks2} and \ref{fig:Real:clocks1}. Such visualization is only approximate, as hinged on the assumption of the joint posterior's elliptic shape; a detailed description of the algorithm for obtaining quantile ellipsoids, \texttt{EllipsoidQuantile[$\cdot$]}, run under Mathematica 12.3, can be found in  \url{https://reference.wolfram.com/language/MultivariateStatistics/ref/EllipsoidQuantile.html}. As seen in the figures, the posterior uncertainty can vary distinctly across the growth rate cycles' points for both identified types of economic fluctuations. Therefore, it appears informative to turn to a probabilistic approach to marking the growth rate cycles' phases, through calculation of the posterior probability of each point belonging to a given quadrant, $Quad_i$, $i=1,...,4$, that is $\Pr\left\{(\Delta K_t, K_t) \in Quad_i |y \right\}$ and $\Pr\left\{(\Delta J_t, J_t) \in Quad_i |y \right\}$, at each $t=1,..., n$. The probabilities presented in Figure \ref{fig:Real:cyclesProbs} illustrate neatly the succession of consecutive phases, indicating the uncertainty as to which stage of a cycle the economy is in at each point in time.



\vspace{-0.12 cm}

\vspace{-0.025 cm}

\vspace{-0.05 cm}

\vspace{-0.25 cm}

\vspace{-0.2 cm}
\section{Discussion and conclusions}

Due to its state space innovation term representation, the stochastic cycle model introduced in this paper belongs to the class of SSOE models, commonly entertained by forecasters. Note that while our basic model specification presented in this paper focuses only on the cyclic fluctuations, it is straightforward to extend it further with typical components such as the level, trend or seasonal pattern. With respect to other approaches to modelling business cycles, the novelty of our model's design resides in its specification being based on trigonometric functions, yet not in a deterministic fashion (affecting the unconditional mean, and thus resulting in a non-stationary in mean process), but with allowing for stochastically governed shifts in the amplitude and phase. As a result, we obtain a covariance stationary model, with its unconditional expectation at zero, time-varying conditional means (driven by trigonometric components endowed with stochastic properties), and finally, a cyclic autocorrelation structure. In that regard, our model remains akin to the one formulated previously in \cite{Harvey_Trimbur_2003} and \cite{trimbur_06}, although both differ distinctly in their specifications.

The model developed in this paper features some relevant theoretical properties  pertaining to its moment structure (proofs presented in Section C of the Supplementary Material). Moreover, it allows to distill from the data individual cyclic components, $C_{j,t}$, $j=1,...,k$, of different (and time-variable) lengths, and corresponding to the frequencies, $\lambda_j$s. Consequently, the interpretability of the isolated components is warranted, in correspondence with typical economic cyclic fluctuations discussed in the literature, and with also capturing the asymmetry of their phases further adding to the model's potential, as evidenced by the empirical study presented above.

For the estimation of our model, we resorted to and developed a Bayesian framework. The Bayesian approach provides a formal and mostly valid setup for handling the uncertainty inherent to model parameters, which appears particularly crucial for models of complex and nonlinear structures, such as the one considered in the present work. Consequently, through basic operations on the simulated posterior sample, a fully probabilistic inference can be conducted on the isolated cyclic components, which enables one to mark economic fluctuations phases through relevant posterior probabilities related to corresponding points on business cycle clocks.

Employing the Bayesian setup necessitated development of relevant posterior simulation methods. The Markov Chain Monte Carlo procedure designed in our paper hinges on the Random Walk Metropolis-Hastings algorithm, which turned out to require here long and numerous burn-in passes to warrant convergence and to offset strong autocorrelations of the resulting chain. Moreover, to attain a reasonable acceptance rate for a given data set, it may also require the researcher to make case-specific adjustments of the algorithm's settings, such as modifying the number of degrees of freedom or, potentially, the manner in which the proposal distribution's covariance matrix is calibrated along the way. Overall, it may appear desirable to undertake further efforts to enhance the algorithm designed in our work, or to seek other, alternative solutions for more efficient sampling from the posterior.

Conceivably, our sampling model leaves some room for improvement and extensions, too. In its present specification, the amplitude and phase shifts changes are driven by processes $A_t$ and $P_t$, respectively, which are \textit{common} to all (rather than specific to each) cyclic components related to various frequencies $\lambda_j$, $j=1,...,k$. However, it may emerge more adequate to individualize both $A_t$ and $P_t$ for each frequency, and supplant $A_t$ and $P_t$ by some $A_{j,t}$ and $P_{j,t}$, with each of the individual processes defined separately. Also, our current model structure admits straightforward incorporation of additional exogenous variables into the equations of both $A_t$ and $P_t$, without detriment to the theoretical properties of the model. Furthermore, yet on a different, more 'structural' level, our present model specification could be generalised into an unobserved components model framework, similar to the one considered in \cite{West1995BayesianII}, where both the amplitude and phase shifts processes are driven by different error terms, also separate from the one governing the observables. Such an approach would 'disentangle' the stochastic changes in the amplitude from the ones pertaining to the phase shifts, with both $A_t$ and $P_t$ now constituting some latent variables. It is expected at the moment that under such a generalised framework, relevant theorems about the model's properties could be proved, much in the spirit of our current research. Obviously, the model's structure would gain plenty in terms of its flexibility, which could prove empirically valid in many cases. However, transforming the model into the latent variables setup would necessitate development of adequate numerical methods to sample from the posterior distribution. Conceivably, a more flexible structure of the unobserved components model could greatly facilitate the convergence of the sampler, thus allowing for a more 'streamlined' and shorter burn-in stage. Moreover, one can also expect that in such a model the numerical issue of an unacceptably high cumulative rounding error of multiple, recursive computation of nested sine functions would be at least alleviated, if present at all, since the likelihood function would no longer require such numerically complex and recursive calculations. This, in turn, would allow modelling longer time series than the one considered in this paper, and enable more precise inference about longer economic cycles.

Eventually, it is mostly desirable that the predictive performance of our model be evaluated through a series of ex post forecasting exercises. For the sake of prediction, it is crucial to ensure that a given model be able to extrapolate the cyclical patterns beyond the sample. This would require pseudo-cyclicality of the conditional expectation, given the past of the process, i.e. $\hat y_{t+h|t}\equiv E(y_{t+h}|\mathcal{F}_{t})$. In Section C of the Supplementary Material, we formulate and prove Theorem C.1, from which it follows that in our model the conditional mean is, indeed, pseudo-cyclical with respect to the forecast horizon $h=1, 2, ...$, with some exponential damping factor. Moreover, since the model proposed in this paper enjoys covariance stationarity, non-explosive long-term predictions are warranted. We also note that the formulae displayed in Theorem C.1 are similar to the ones obtained for the 'classic' stochastic cycle model with an $\text{ARMA}(2,1)$ structure; see Theorem 3.1 in \cite{Pelagatti_16}.








\bibliographystyle{apa}
\bibliography{biblioOK}
\end{document}